\documentclass[11pt]{article}
\usepackage[dvipdfmx]{graphicx}   
\usepackage{amssymb}
\usepackage{array} 
\usepackage[authoryear]{natbib}
\usepackage{bibentry}
\usepackage{amsmath}
\usepackage{float}
\usepackage{verbatim}
\usepackage{color}
\usepackage{makeidx}
\usepackage{bm}
\usepackage{appendix}
\usepackage{lscape}
\usepackage{subcaption}

\begin{document}
\title{Stochastic Volatility in Mean: Efficient Analysis \\ by a Generalized Mixture Sampler}
\author{\textsc{Daichi Hiraki} \\
\textit{Graduate School of Economics, University of Tokyo, Tokyo 113-0033, Japan} \\
\texttt{hdaichi397@gmail.com}
\and \textsc{Siddhartha Chib} \\
\textit{Olin School of Business, Washington University, St Louis, USA} \\
\texttt{chib@wustl.edu} 
\and \textsc{Yasuhiro Omori} \\
\textit{Faculty of Economics, University of Tokyo, Tokyo 113-0033, Japan}\\
\texttt{omori@e.u-tokyo.ac.jp} \\
}
\date{November 2024}
\maketitle

\begin{abstract}
In this paper we consider the simulation-based Bayesian analysis of stochastic volatility in mean (SVM) models. 
Extending the highly efficient Markov chain Monte Carlo mixture sampler for the SV model proposed in \cite{KimShephardChib(98)} and \cite{OmoriChibShephardNakajima(07)}, we develop an accurate approximation of the non-central chi-squared distribution as a mixture of thirty normal distributions. Under this mixture representation, we sample the parameters and latent volatilities in one block. We also detail a correction of the small approximation error by using additional Metropolis-Hastings steps. The proposed method is extended to the SVM model with leverage. The methodology and models are applied to excess holding yields and S\&P500 returns in empirical studies, and the SVM models are shown to outperform other volatility models based on marginal likelihoods.
\end{abstract}
{\bf JEL classification}: C11, C15, C32, C58
\\
{\bf Keywords}: Excess Holding Yield; Markov chain Monte Carlo; Mixture Sampler; Risk Premium; Stochastic Volatility in Mean

\newpage
\section{Introduction}
In financial time series, volatility clustering, the phenomenon of persistent volatility, is well known to exist. One way to model time-varying volatility, or volatility clustering, is by using the stochastic volatility (SV) model of \cite{Taylor(08)}. In the simplest version of this model, the standard deviation of the outcome is given by an exponential transformation of an unobserved log-variance variable $h_t$, where $h_t$ in turn is modeled by a stationary first-order autoregressive process. The model in this basic form can be viewed as a state-space model in which the measurement equation is nonlinear in the latent variance $h_t$. A significant variant of this standard SV model is one in which the standard deviation of the outcome $\exp(h_t/2)$ appears as a predictor variable in the mean of the measurement equation. This is called the SV in mean model (SVM) and is similar in spirit
to the ARCH in mean (ARCH-M) model introduced by \cite{EngelLilienRobins(87)}. 
Like the standard SV model, the SVM model has also been used in various fields, including
macroeconomics and finance (see \cite{KooopmanHol(02)}, \cite{BerumentYalcinYildirim(09)}, \cite{MumtazZanetti(13)}, \cite{CrossHouKoopPoon(23)}).

In the Markov chain Monte Carlo (MCMC) estimation of model parameters for the SV-type models, it is often observed that sampling one latent variable conditional on all the other latent variables and parameters, which is referred to as the single-move sampler, is inefficient in the sense that MCMC draws are highly autocorrelated. To address this issue, \cite{KimShephardChib(98)} introduced the mixture sampler as a highly efficient Bayesian estimation method for the standard SV model. This approach was extended to SV models with jumps and fat-tailed errors in \cite{ChibNardariShephard(02)} and to SV models with leverage in \cite{OmoriChibShephardNakajima(07)}.

In the existing literature, the SVM model without leverage has been estimated by the multi-move (block) sampler (\cite{ShephardPitt(97)} and \cite{OmoriWatanabe(08)}), for example, \cite{AbantoMigonLachos(11)}, \cite{AbantoMigonLachos(12)}, and \cite{LeaoAbantoChen(17)}, and by other similar approaches, for example, \cite{Chan(17)}, \cite{AbantoRodriguezGarrafa(21)}, and \cite{AbantoRodriguezHernan(23)}. There is no known mixture sampler approach for SVM models with leverage.  In this paper, we develop efficient MCMC based algorithms for SVM models, with and without leverage, that are based on accurate representations of these models in terms of mixtures of conditionally Gaussian linear state-space models, just as in the approach of \cite{KimShephardChib(98)}. However, due to the $\beta\exp(h_t/2)$ term in the mean equation, instead of characterizing the distribution of a central chi-squared distribution with one degrees of freedom in terms of mixtures of normal distributions, a mixture representation to the distribution of a non-central chi-squared distribution with one degrees of freedom is needed, in which the non-centrality parameter is $\beta^2$. We show that the latter distribution has an infinite series expansion. Our estimation approach uses a truncated version of this series expansion to develop a highly efficient fitting algorithm. It can be viewed as a generalized mixture sampler. Furthermore, the small approximation error due to the truncation can be corrected by a data augmentation method by incorporating a pseudo-target probability density whose marginal probability density is the exact conditional posterior density. 

We apply our proposed method to excess holding yields and S\&P500 returns data, and show that SVM models outperform other SV models based on marginal likelihoods.
%
The rest of this paper is organized as follows.
Section 2 introduces the SVM model and describes the novel mixture sampler as an efficient sampling method for such models. 
The MCMC simulation and the particle filter are described in Section 3. 
Section 4 illustrates the performance of this sampling method using the simulated data for several cases. Section 5 further extends the SVM model to incorporate the leverage effect. Finally, in Section 6, we apply our proposed SVM model to financial data and perform a model comparison. Conclusion and remarks are given in Section 7.

\section{Stochastic volatility in mean model}
\subsection{SVM model}
We define the stochastic volatility in mean (SVM) model as follows:
\begin{align} 
        y_t &= \beta \exp(h_t/2) + \epsilon_t \exp(h_t/2), \quad
        t = 1, ..., n, \label{SVM obs} \\ 
        h_{t+1} &= \mu + \phi(h_t - \mu) + \eta_t, \quad
        t = 1, ..., n-1, \label{SVM state} \\
        &\begin{pmatrix}
            \epsilon_t \\
            \eta_t
        \end{pmatrix}
        \overset{\text{i.i.d.}}{\sim} N(0, \Sigma), \quad
        \Sigma = 
        \begin{pmatrix}
            1 & 0 \\
            0 & \sigma^2
        \end{pmatrix},  \label{eq:error_without_leverage}\\
        &h_1 \sim N \left(\mu, \frac{\sigma^2}{1-\phi^2} \right),
\end{align}
where $N(m,S)$ denotes normal distribution with mean vector $m$ and covariance matrix $S$, $\theta = (\mu, \phi, \sigma^2, \beta)$ is a model parameter vector of interest, and $h = (h_1, \dots, h_n)'$ is the logarithm of the latent volatility vector. We use the standard deviation $\exp(h_t/2)$ in the mean equation, rather than the variance $\exp(h_t)$, to match the units of the outcome variable. 
For $\beta \neq 0$, we denote it as the stochastic volatility in mean (SVM) model. The standard stochastic volatility (SV) model is obtained as a special case with $\beta =0$.  
For $\theta$, we assume the prior distribution
\begin{align*}
    &\mu \sim N(\mu_0, \sigma_0^2), \quad
    \frac{\phi+1}{2} \sim Beta(a,b), \\
    &\sigma^2 \sim IG \left( \frac{n_0}{2}, \frac{s_0}{2} \right),  \quad
    \beta \sim N(b_0, B_0).
\end{align*}
where $Beta(a,b)$ denotes beta distribution with parameters $(a,b)$ and $IG(a,b)$ denotes inverse gamma distribution with parameters $(a,b)$ whose probability density function is
\begin{align*}
& \pi(x|a,b) \propto x^{-(a+1)}\exp(-bx),\quad x >0, \quad a,b>0. 
\end{align*}
We let $f(y, h|\theta)$ and $\pi(\theta)$ denote the probability density function of $(y, h)$ given $\theta$ and the prior probability density function of $\theta$ where $y \equiv (y_1, \dots, y_n)'$. The posterior density function of $(h, \theta)$ is given in Appendix \ref{appendix:correction}.
\vspace{1mm}\\

\noindent
{\it Remark}. It is straightforward to include the constant term in the measurement equation. However, noting that $\beta \times \exp(h_t/2) \approx  \beta \times (1 + h_t/2)$, it is often confounded with $\beta$ and therefore omitted in this paper. 
\subsection{Transformation of the measurement equation}
To sample $h$ from its conditional distribution, we transform Equation (\ref{SVM obs}) as below:
\begin{equation}
    y_t^* = h_t + \epsilon_t^*, \quad y_t^* = \log(y_t^2), \quad \epsilon_t^* = \log(\beta + \epsilon_t)^2,
    \label{transform}
\end{equation}
Since $(\beta + \epsilon_t) \sim N(\beta, 1)$, its square $(\beta + \epsilon_t)^2 \sim \chi_1^2(\beta^2)$ where $\chi_1^2(\beta^2)$ denotes the non-central chi-square distribution with the non-centrality parameter $\beta^2$ and one degrees-of-freedom.
The special case with $\beta=0$ is considered in \cite{KimShephardChib(98)} and \cite{OmoriChibShephardNakajima(07)}, 
who introduced the idea of accurately approximating the probability of the logarithm of the central chi-square distribution with one degrees of freedom, $\log \chi_1^2(0)$,
\begin{eqnarray*}
    f(\epsilon_t^*) = \frac{1}{\sqrt{2\pi}} \exp \left( \frac{\epsilon_t^* - \exp(\epsilon_t^*)}{2} \right), \quad -\infty <\epsilon_t^* < \infty,
\end{eqnarray*}
by a mixture of normal distributions. Below, we elaborate a highly accurate approximation of the distribution of $\epsilon_t^*$ given $\beta \neq 0$ by the mixture of normal distributions. 
Let $p(x;\nu, \lambda)$ be the probability density function of $\chi_\nu^2(\lambda)$. 
It can be expressed as an infinite mixture of central $\chi^2$ probability density functions (see, e.g. \cite{JohnsonKotzBalakrishnan(95)}):
\begin{eqnarray*}
    p(x;\nu, \lambda) &=& \sum_{j=0}^{\infty} \left\{ \frac{\left(\frac{\lambda}{2}\right)^j}{j!} \exp\left(-\frac{\lambda}{2}\right) \right\} p(x;\nu+2j, 0),
    \label{chi2_pdf}
\\
    && p(x; \nu + 2j, 0) = \frac{x^{\frac{\nu}{2}+j-1}}{2^{\frac{\nu}{2}+j} \Gamma(\frac{\nu}{2}+j)} \exp\left(-\frac{x}{2}\right).
    \nonumber
\end{eqnarray*}
Setting  $\nu=1$ and noting that
\begin{equation*}
    p(x;1+2j, 0) = \frac{x^j \Gamma\left(\frac{1}{2}\right)}{2^j \Gamma\left(\frac{1}{2}+j\right)} \times p(x; 1, 0),
\end{equation*}
we obtain the expression
\begin{eqnarray}
    p(x;1,\lambda) &=& \sum_{j=0}^{\infty} \left\{ \frac{\left(\frac{\lambda}{2}\right)^j}{j!} \exp\left(-\frac{\lambda}{2}\right) \right\} 
    \label{chi2_pdf_df=1}
    \frac{x^j \Gamma\left(\frac{1}{2}\right)}{2^j \Gamma\left(\frac{1}{2}+j\right)} \times p(x; 1, 0).
\end{eqnarray}
Let $f(u;\lambda)$ denote the probability density function of $U \sim \log \chi_1^2(\lambda)$.
Using (\ref{chi2_pdf_df=1}),  it follows that
\begin{equation}
    f(u;\lambda) = \sum_{j=0}^{\infty} \frac{\left(\frac{\lambda}{2}\right)^j}{j!} \exp\left(-\frac{\lambda}{2}\right) \frac{\exp(uj) \Gamma\left(\frac{1}{2}\right)}{2^j \Gamma\left(\frac{1}{2}+j\right)} \times f(u;0).
    \label{approx_chi2}
\end{equation}
As in \cite{OmoriChibShephardNakajima(07)}, we consider the mixture of ten normal distributions to approximate $f(u;0)$,  the probability density function of $\log \chi_1^2(0)$,
\begin{equation}
    f(u;0) \approx \sum_{i=1}^{K} p_i v_i^{-1} \phi \left( \frac{u-m_i}{v_i} \right), \quad K=10,
    \label{approx_omori}
\end{equation}
where $\phi(\cdot)$ denotes the probability density function of the standard normal distribution. The values of $(p_i, m_i, v_i^2)$ are taken from \cite{OmoriChibShephardNakajima(07)} and are reproduced in Table \ref{table:approx_chi2}. The columns labeled $a_i$ and $b_i$ will be used when we consider the model with leverage in Section \ref{sec:approx_SVM_leverage}.
\begin{table}[H]
    \small
  \centering
  \begin{tabular}{|c|rrrrr|}
    \hline 
    $i$ & $p_i$  & $m_i$ & $v_i^2$ & $a_i$ & $b_i$ \\
    \hline
    1 & 0.00609 & 1.92677 & 0.11265 & 1.01418 & 0.50710 \\
    2 & 0.04775 & 1.34744 & 0.17788 & 1.02248 & 0.51124 \\
    3 & 0.13057 & 0.73504 & 0.26768 & 1.03403 & 0.51701 \\
    4 & 0.20674 & 0.02266 & 0.40611 & 1.05207 & 0.52604 \\
    5 & 0.22715 & -0.85173 & 0.62699 & 1.08153 & 0.54076 \\
    6 & 0.18842 & -1.97278 & 0.98583 & 1.13114 & 0.56557 \\
    7 & 0.12047 & -3.46788 & 1.57469 & 1.21754 & 0.60877 \\
    8 & 0.05591 & -5.55246 & 2.54498 & 1.37454 & 0.68728 \\
    9 & 0.01575 & -8.68384 & 4.16591 & 1.68327 & 0.84163 \\
    10 & 0.00115 & -14.65000 & 7.33342 & 2.50097 & 1.25049 \\
    \hline
  \end{tabular}
  \caption{Selection of $(p_i, m_i, v_i^2, a_i, b_i)$ introduced in \cite{OmoriChibShephardNakajima(07)}.}
  \label{table:approx_chi2}
  \normalsize
\end{table}
By substituting Equation (\ref{approx_omori}) to Equation (\ref{approx_chi2}), we obtain
\begin{eqnarray}
    f(u; \lambda) &\approx &\sum_{j=0}^{\infty} \frac{\left(\frac{\lambda}{2}\right)^j}{j!} \exp\left(-\frac{\lambda}{2}\right) \frac{\exp(uj) \Gamma\left(\frac{1}{2}\right)}{2^j \Gamma\left(\frac{1}{2}+j\right)} \sum_{i=1}^{K} p_i v_i^{-1} \phi \left( \frac{u-m_i}{v_i} \right) 
    \nonumber \\
    &= & \sum_{i=1}^{K}  \sum_{j=0}^{\infty} p_i \exp \left(-\frac{\lambda}{2} \right) \frac{\Gamma\left(\frac{1}{2}\right)}{2^j j! \Gamma\left(\frac{1}{2} + j\right)} \left(\frac{\lambda}{2}\right)^j \exp \left( m_i j + \frac{j^2 v_i^2}{2} \right) 
    \nonumber \\
    && \hspace{3cm} \times \frac{1}{\sqrt{2 \pi v_i^2}} \exp \left\{ -\frac{(u-(m_i+j v_i^2))^2}{2 v_i^2} \right\} 
    \nonumber \\
    &=& \sum_{i=1}^{K}  \sum_{j=0}^{\infty} p_i \exp \left(-\frac{\lambda}{2} + m_i j + \frac{j^2 v_i^2}{2} \right) \frac{\Gamma\left(\frac{1}{2}\right)}{2^j j! \Gamma\left(\frac{1}{2} + j\right)} \left(\frac{\lambda}{2}\right)^j 
    \nonumber \\
    && \hspace{3cm} \times v_i^{-1} \phi \left( \frac{u-(m_i+j v_i^2)}{v_i} \right) 
    \nonumber\\
    & \approx & \sum_{i=1}^{K} \sum_{j=0}^{J} \Tilde{p}_{i,j} v_i^{-1} \phi \left( \frac{u-\Tilde{m}_{i,j}}{v_i} \right), 
    \label{approx_general_noncentral_chi2}
\end{eqnarray}
where 
\begin{equation*}
    \Tilde{p}_{i,j} = \frac{w_{i,j}}{\sum_{i=1}^{K} \sum_{j=0}^{J} w_{i,j}}, \quad
    w_{i,j} = p_i \exp \left(-\frac{\lambda}{2} + m_i j + \frac{j^2 v_i^2}{2} \right) \frac{\Gamma\left(\frac{1}{2}\right)}{2^j j! \Gamma\left(\frac{1}{2} + j\right)} \left(\frac{\lambda}{2}\right)^j,
\end{equation*}
and $\Tilde{m}_{i,j} = m_i+j v_i^2$. In the last equality, we truncate the summation at $j=J$ and normalize $\Tilde{p}_{i,j}$ to ensure that $\sum_{i=1}^{K} \sum_{j=0}^{J}  \Tilde{p}_{i,j} = 1$.

\begin{figure}[H]
    \centering
    \begin{subfigure}{0.325\linewidth}
        \centering
        \includegraphics[width=\linewidth]{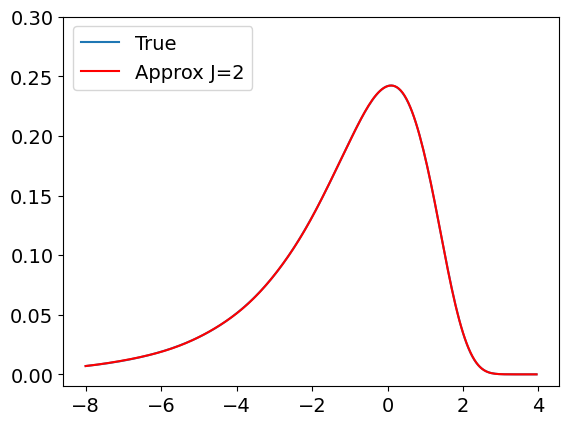}
        \caption{$\beta = 0.3$}
    \end{subfigure}
    \hfill
    \begin{subfigure}{0.325\linewidth}
        \centering
        \includegraphics[width=\linewidth]{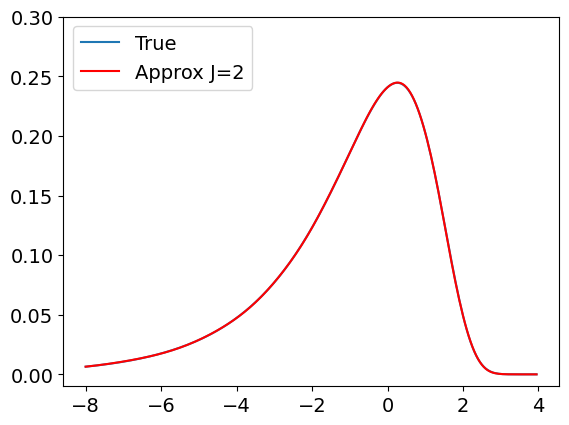}
        \caption{$\beta = 0.5$}
    \end{subfigure}
    \hfill
    \begin{subfigure}{0.325\linewidth}
        \centering
        \includegraphics[width=\linewidth]{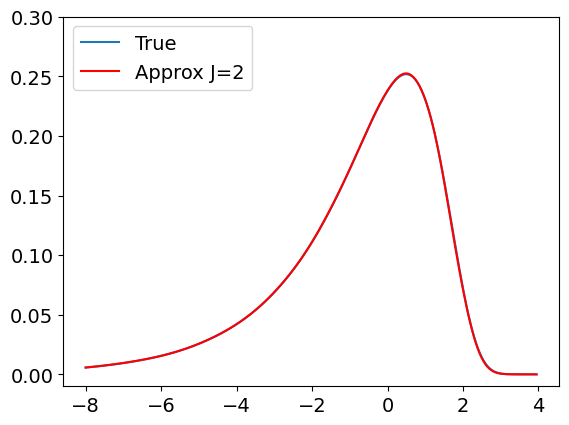}
        \caption{$\beta = 0.7$}
    \end{subfigure}
    \caption{True and approximation densities of $\log \chi_1^2(\beta^2)$ for $\beta = 0.3,0.5$ and $0.7$.}
    \label{fig:approx}
\end{figure}

\begin{figure}[H]
    \centering
    \begin{subfigure}{0.325\linewidth}
        \centering
        \includegraphics[width=\linewidth]{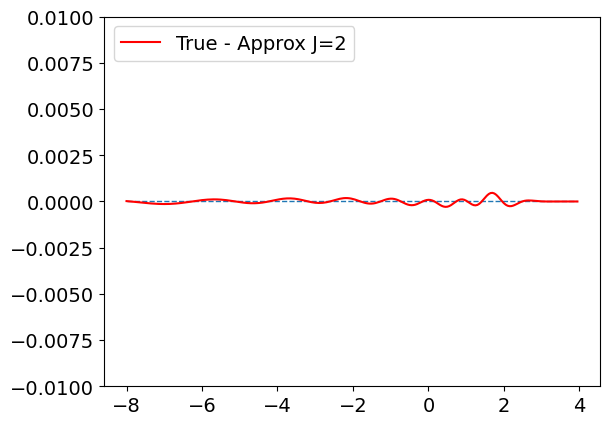}
        \caption{$\beta = 0.3$}
    \end{subfigure}
    \hfill
    \begin{subfigure}{0.325\linewidth}
        \centering
        \includegraphics[width=\linewidth]{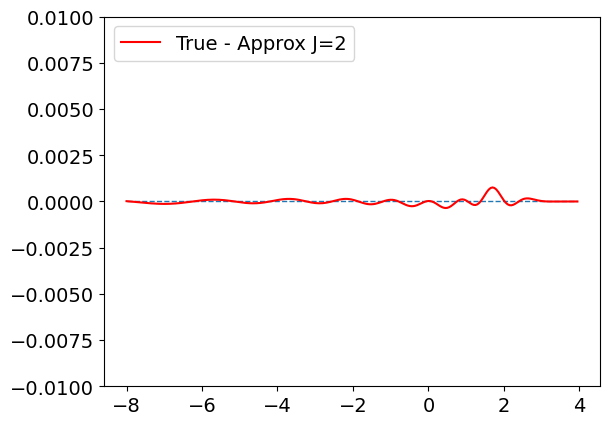}
        \caption{$\beta = 0.5$}
    \end{subfigure}
    \hfill
    \begin{subfigure}{0.325\linewidth}
        \centering
        \includegraphics[width=\linewidth]{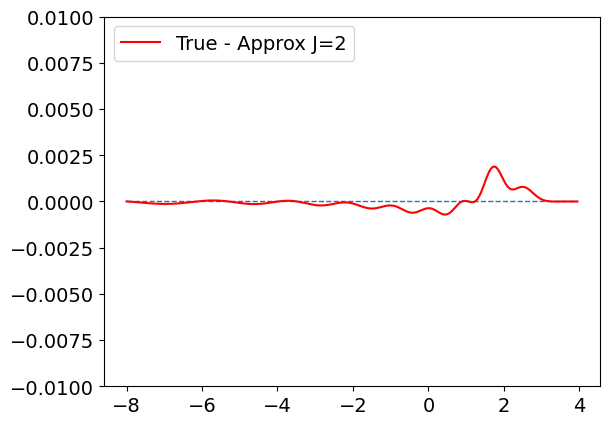}
        \caption{$\beta = 0.7$}
    \end{subfigure}
    \caption{Differences between the true density of $\log \chi_1^2(\beta^2)$ and the approximate densities for $\beta = 0.3,0.5$ and $0.7$.}
    \label{fig:diff}
\end{figure}

This expression implies that the probability density function of $\log \chi_1^2(\lambda)$ is approximated by the mixture of $K(J+1)$ normal distributions.
Especially, when $\lambda = 0$ and $J=0$, the approximation (\ref{approx_general_noncentral_chi2}) reduces to (\ref{approx_omori}). 
The extent of the impact on the approximation is based on the value $w_{i,j}$ which includes $\lambda=\beta^2$.
The coefficient $\beta$ of volatility $\exp(h_t/2)$ is estimated to be less than one in past empirical studies.
When $\lambda = \beta^2$ is less than 1.0, $J = 1$ or $2$ makes $\sum_{i=1}^{K} \sum_{j=0}^{J} w_{i,j}$ more than 0.9.
For $J=2$, Figures \ref{fig:approx} and \ref{fig:diff} show the true and approximate densities of $\log \chi_1^2(\beta^2)$ for $\beta = 0.3, 0.5$ and $0.7$ (equivalently, $\beta = -0.3, -0.5$ and $-0.7$) and the difference between the two densities, respectively.
Since these differences are quite small and the approximation  almost overlaps the true probability density of $\log \chi_1^2(\beta^2)$ even for $\beta=0.7$, we employ $J = 2$ in this paper. That is, we approximate the probability density of $\epsilon_t^*|\beta \sim \log \chi_1^2(\beta^2)$ by

\begin{equation}
    f(\epsilon_t^*|\beta) \approx \sum_{i=1}^{10} \sum_{j=0}^{2} \Tilde{p}_{i,j} v_i^{-1} \phi \left( \frac{u-\Tilde{m}_{i,j}}{v_i} \right),
    \label{approx_noncentral_chi2}
\end{equation}
where
\begin{equation*}
    \Tilde{p}_{i,j} = \frac{p_i \exp \left( m_i j + \frac{j^2 v_i^2}{2} \right) \frac{1}{2^j j! \Gamma(1/2 + j)} \left(\frac{\beta^2}{2}\right)^j}{\sum_{i=1}^{10} \sum_{j=0}^{2} p_i \exp \left( m_i j + \frac{j^2 v_i^2}{2} \right) \frac{1}{2^j j! \Gamma(1/2 + j)} \left(\frac{\beta^2}{2}\right)^j}, \quad \Tilde{m}_{i,j} = m_i+j v_i^2.
\end{equation*}

\noindent
Let $s_t = (s_{1t}, s_{2t}) \in \{ (i,j) | i=1,...,K, j=0,...,J \}$ denote the component of the mixture of the normal densities in (\ref{approx_noncentral_chi2}) at time $t$.
Given $s_t = (i,j)$, we have $\epsilon_t^* | s_t=(i,j) \sim N(\Tilde{m}_{i,j}, v_i^2)$ and we see that the SVM model can be approximated by the linear Gaussian state space form
\begin{align}
    y_t^* &= \Tilde{m}_{s_{1t}, s_{2t}} + h_t + (v_{s_{1t}}, 0) z_t, \label{approx_linaer_start} \\
    h_{t+1} &= \mu(1-\phi) + \phi h_t + (0, \sigma) z_t, \nonumber \\
    &t = 1,...,n-1, \quad h_1 \sim N\left( \mu, \frac{\sigma^2}{1-\phi^2} \right), \quad
    |\phi| < 1, \label{approx_linaer_end} \\
    &z_t = (z_{1t}, z_{2t})' \sim N(0, I_2), \nonumber
\end{align}
where $y_t^* = \log(y_t^2)$.
In the following sections, $\Tilde{m}_{s_{1t}, s_{2t}}$ and $\Tilde{p}_{s_{1t}, s_{2t}}$ are abbreviated as $\Tilde{m}_{s_t}$ and $\Tilde{p}_{s_t}$, and we write $v_{s_t}^2$ instead of $v_{s_{1t}}^2$.
\section{MCMC simulation and associated particle filter}
\subsection{MCMC algorithm}
\textbf{Algorithm 1 (Generalized mixture sampler, GMS).}
Let us denote $\theta = (\alpha,\beta)$ where $\alpha = (\mu,\phi,\sigma^2)$.
The Markov chain Monte Carlo simulation is implemented in four blocks:

\begin{itemize}
    \item[1.] Initialize $h$ and $\theta=(\alpha,\beta)$.
    \item[2.] Generate $\beta|\alpha,h, y \sim \pi(\beta|\alpha,h, y)$.
    \item[3.] Generate $(\alpha,h)|\beta, y \sim \pi(\alpha,h|\beta, y)$.
    \item[4.] Go to Step 2.
\end{itemize}
\subsubsection*{Step 2. Generation of $\beta|\alpha,h,y$}
The conditional posterior distribution of $\beta$ is normal with mean $b_1$ and variance $B_1$ where
\begin{eqnarray*}
b_1 = B_1 \left(X'\Omega^{-1}y + B_0^{-1}b_0\right), \quad 
B_1^{-1} = X'\Omega^{-1}X + B_0^{-1},
\end{eqnarray*}
and
\begin{eqnarray*}
X = \left(\begin{array}{c}  \exp(h_1/2) \\  \exp(h_2/2) \\ \vdots  \\  \exp(h_n/2)  \end{array}  \right),
\quad
\Omega = \mbox{diag}\left(\exp(h_1),\exp(h_2),\cdots, \exp(h_n)\right).
\end{eqnarray*}
Thus we generate $\beta \sim N(b_1,B_1).$
\subsubsection*{Step 3. Generation of $(\alpha,h)|\beta,y$}
\label{generate h}
As discussed in the previous section, we sample $h$ using the mixture sampler using the mixture of normal distributions. Since our approximation is highly accurate, we can use this mixture approximation directly in Step 3, as in Step 2 of Algorithm 1 in \cite{ChibNardariShephard(02)}. However, one can remove the small approximation error with an additional MH step, as detailed in Algorithm 2, GMS with MH algorithm (GMH), given in the Appendix \ref{appendix:correction}, but due to the fact that the tailored mixture very closely fits the non-central chi-squared distribution, this additional step would be rarely necessary. 

Let $f_N(\cdot|m, s^2)$ denote the probability density of $N(m, s^2)$, and let $\pi(\alpha)$ denote the prior density of $\alpha$.
Define our target density in Step 3 as 
\begin{align*}
    \pi^*(\alpha,h,s|\beta,y) &= \pi^*(\alpha,h|\beta,s,y) \times q(s),\\
         &= \pi^*(h|\alpha,\beta,s,y)\pi^*(\alpha|\beta,s,y) \times q(s),\\
     & q(s) = \prod_{t=1}^n \Tilde{p}_{s_t}, \quad y_t^* = \log(y_t^2),
\end{align*}
where 
\begin{align*}
    \pi^*(h|\alpha, s, \beta, y^*) &= \frac{\prod_{t=1}^n g(y_t^*| h_t, \alpha,\beta, s_t)}{m(y^*|\alpha,s,\beta)} \times \prod_{t=1}^{n-1}f_N(h_{t+1}|\mu(1-\phi)+\phi h_t,\sigma^2) \times f_N\left( h_1 \bigg| \mu, \frac{\sigma^2}{1-\phi^2} \right).\\
    \pi^*(\alpha|s,\beta,y^*) & \propto  m(y^*|\alpha,s,\beta) \pi(\alpha),\\
    & g(y_t^*| h_t, \alpha, \beta, s_t) = 
    f_N(y_t^*|\Tilde{m}_{s_t} + h_t, v_{s_t}^2), \quad t=1,\ldots,n, 
\end{align*}
and  $\Tilde{m}_{s_t} = \Tilde{m}_{s_{1t}, s_{2t}}$ and $\Tilde{p}_{s_t} = \Tilde{p}_{s_{1t}, s_{2t}}$ are defined in (\ref{approx_noncentral_chi2}) and $(p_{s_{1t}}, m_{s_{1t}}, v_{s_{1t}}^2)$ are given in Table \ref{table:approx_chi2}.
Note that $m(y^*|\alpha,s,\beta)$ is a normalizing constant for $\pi^*(h|\alpha, s, \beta, y^*)$ and is evaluated using the Kalman filter algorithm. Only $\Tilde{p}_{s_t}$, which depends on $\beta$, needs to be updated according to the formula in (\ref{approx_noncentral_chi2}) before sampling.
Note that the target density $\pi^*(\alpha,h|\beta,y)$ approximates the true conditional density $\pi(\alpha,h|\beta,y)$ accurately. We generate the sample $(\alpha, h, s)$ in two steps. 
\begin{itemize}
    \item[(a)] Generate $s \sim q(s|h,\alpha, \beta, y^*)$ where
    \begin{align*}
    &q(s|h,\alpha,\beta, y^*) = \prod_{t=1}^n \frac{\Tilde{p}_{s_t} g(y_t^*| h_t, \alpha,\beta, s_t)}{\sum_{i=1}^{10}\sum_{j=0}^2 \Tilde{p}_{i,j} g(y_t^*| h_t, \alpha,\beta, s_t=(i,j))}.
    \end{align*}

    \item[(b)] Generate $(\alpha,h)\sim \pi^*(\alpha,h|s,\beta,y)$
        \begin{itemize}
        \item[(i)]  Generate $\alpha \sim \pi^*(\alpha|s,\beta,y^*)$. 
        We first transform $\alpha$ to $\vartheta = (\mu, \log\{ (1+\phi)/(1-\phi) \}, \log \sigma^2)$ to remove parameter constraints and perform the Metropolis-Hastings (MH) algorithm \citep{ChibGreenberg95} to sample from the conditional posterior distribution with density $\pi^*(\vartheta|s,\beta,y) = \pi^*(\alpha|s, \beta,y) \times |d\alpha / d\vartheta|$ where $|d\alpha / d\vartheta|$ is the Jacobian of the transformation. 
        Compute the posterior mode $\hat{\vartheta}$ and define $\vartheta_*$ and $\Sigma_*$ as
        \begin{equation*}
        \vartheta_* = \hat{\vartheta}, \quad
        \Sigma_*^{-1} = -\frac{\partial^2 \log \pi^*(\vartheta|s,\beta, y)}{\partial \vartheta \partial \vartheta'} \bigg|_{\vartheta = \hat{\vartheta}}.
        \end{equation*}
    
        Given the current value $\vartheta$, generate a candidate $\vartheta^\dag$ from the distribution $N(\vartheta_*, \Sigma_*)$ and accept it with probability
        \begin{equation*}
        \alpha(\vartheta, \vartheta^\dag|s,\beta, y) = \min \left\{1, \frac{\pi^*(\vartheta^\dag|s, \beta, y) f_N(\vartheta|\vartheta_*, \Sigma_*)}{\pi^*(\vartheta|s, \beta,y) f_N(\vartheta^\dag|\vartheta_*, \Sigma_*)} \right\},
        \end{equation*}
        where $f_N(\cdot|\vartheta_*, \Sigma_*)$ is the probability density of $N(\vartheta_*, \Sigma_*)$. 
        If the candidate $\vartheta^\dag$ is rejected, we take the current value $\vartheta$ as the next draw. When the Hessian matrix is not negative definite, we may take a flat normal proposal $N(\vartheta_*, c_0 I)$ using some large constant $c_0$. 
        \item[(ii)] Generate $h|\alpha, s, \beta, y \sim \pi^*(h|\alpha, s, \beta, y)$. We generate $h = (h_1, ..., h_n)$ using a simulation smoother introduced by \cite{DeShephard(95)} and \cite{DurbinKoopman(02)} for the linear Gaussian state space model as in (\ref{approx_linaer_start})-(\ref{approx_linaer_end}).

    \end{itemize}
\end{itemize}
\bigskip

\subsection{Associated particle filter}
\label{partile filter}
We describe how to compute the likelihood $f(y|\theta)$
\begin{equation*}
    f(y|\theta) = \int f(y, h|\theta) dh,
\end{equation*}
numerically as it is necessary to obtain the marginal likelihood, $f(y) = \int f(y|\theta) \pi(\theta) d\theta$ and Bayes factor for the model comparison. 
The filtering and the associated particle computations are carried out by the auxiliary particle filter (see e.g. \cite{PittShephard(99)}, \cite{OmoriChibShephardNakajima(07)}).
Let us denote $Y_t = (y_1, \dots, y_n)$, and
\begin{align*}
    f(y_t|h_t, \theta) &= \frac{1}{\sqrt{2\pi}} \exp \left[ -\frac{1}{2}h_t - \frac{1}{2} \{ y_t - \beta \exp(h_t/2) \}^2 \exp(-h_t) \right] \\
    f(h_{t+1}|h_t, y_t, \theta) &= \frac{1}{\sqrt{2\pi (1-\rho^2)} \sigma} \exp \left\{ -\frac{(h_{t+1} - \mu_{t+1})^2}{2\sigma^2} \right\}, \\
    \mu_{t+1} &= \mu + \phi(h_t-\mu),
\end{align*}
and consider the importance function for the auxiliary particle filter
\begin{align*}
    q(h_{t+1}, h_t^i| Y_{t+1}, \theta) &\propto  f(y_{t+1}| \mu_{t+1}^i, \theta) f(h_{t+1}|h_t^i, y_t, \theta) \hat{f}(h_t^i|Y_t, \theta) \\
    &\propto f(h_{t+1}|h_t^i, y_t, \theta) q(h_t^i| Y_{t+1}, \theta)
\end{align*}
where
\begin{align*}
    q(h_t^i| Y_{t+1}, \theta) &= \frac{f(y_{t+1}| \mu_{t+1}^i, \theta) \hat{f}(h_t^i|Y_t, \theta)}{\sum_{j=1}^I f(y_{t+1}| \mu_{t+1}^j, \theta) \hat{f}(h_t^j|Y_t, \theta)}, \\
    f(y_{t+1}|\mu_{t+1}^i, \theta) &= \frac{1}{\sqrt{2\pi}} \exp \left[ -\frac{1}{2}\mu_{t+1}^i - \frac{1}{2} \{ y_t -  \beta \exp(h_t^i/2) \}^2 \exp(-\mu_{t+1}^i) \right], \\
    \mu_{t+1}^i &= \mu + \phi(h_t^i-\mu).    
\end{align*}
This leads to the following particle filtering.
\begin{itemize}
    \item[1.] Compute $\hat{f}(y_1|\theta)$ and $\hat{f}(h_1^i| Y_1, \theta) = \pi_1^i$ for $i = 1, \dots, I$. 
    
    \begin{itemize}
        \item[(a)] Generate $h_1^i \sim f(h_1|\theta)$ ($= N(\mu, \sigma^2/(1-\phi^2))$) for $i = 1, \dots, I$.
        
        \item[(b)] Compute
        \begin{align*}
            &\pi_1^i = \frac{w_i}{\sum_{i=1}^I w_i}, \quad
            w_i = f(y_1|h_1, \theta), \quad 
            W_i = F(y_1|h_1, \theta), \\
            &\hat{f}(y_1|\theta) = \overline{w}_1 = \frac{1}{I} \sum_{i=1}^I w_i, \quad
            \hat{F}(y_1|\theta) = \overline{W}_1 = \frac{1}{I} \sum_{i=1}^I W_i,
        \end{align*}
        where $f(y_1|\theta)$ and $F(y_1|\theta)$ are the marginal density function and the marginal distribution function of $y_1$ given $\theta$. Let $t = 1$.
    \end{itemize}

    \item[2.] Compute $\hat{f}(y_{t+1}|\theta)$ and $\hat{f}(h_{t+1}^i| Y_{t+1}, \theta) = \pi_{t+1}^i$ for $i = 1, \dots, I$.
    
    \begin{itemize}
        \item[(a)] Sample $h_t^i \sim q(h_t| Y_t, \theta)$, $i = 1, \dots, I$.

        \item[(b)] Generate $h_{t+1}^i| h_t^i, y_t, \theta \sim f(h_{t+1}| h_t^i, y_t, \theta)$ ($= N(\mu_{t+1}^i, \sigma^2)$) for $i = 1, \dots, I$.  

        \item[(c)] Compute
    \end{itemize}
       \begin{align*}
            &\pi_{t+1}^i = \frac{w_i}{\sum_{i=1}^I w_i}, \quad
            w_i = \frac{ f(y_{t+1}|h_{t+1}^i, \theta) f(h_{t+1}^i| h_t^i, y_t, \theta) \hat{f}(h_t^i| Y_t, \theta) }{ f(h_{t+1}^i| h_t^i, y_t, \theta) q(h_t^i| Y_{t+1}, \theta) } = \frac{ f(y_{t+1}|h_{t+1}^i, \theta) \hat{f}(h_t^i| Y_t, \theta) }{ q(h_t^i| Y_{t+1}, \theta) }, \\ 
            &W_i = \frac{ F(y_{t+1}|h_{t+1}^i, \theta) \hat{f}(h_t^i| Y_t, \theta) }{ q(h_t^i| Y_{t+1}, \theta) }, \\
            &\hat{f}(y_{t+1}| Y_t, \theta) = \overline{w}_{t+1} = \frac{1}{I} \sum_{i=1}^I w_i, \quad
            \hat{F}(y_{t+1}|\theta) = \overline{W}_{t+1} = \frac{1}{I} \sum_{i=1}^I W_i.
        \end{align*}    

        \item[3.] Increment $t$ and go to 2.
\end{itemize}
\noindent
It can be shown that as $I \rightarrow \infty$, $\overline{w}_{t+1} \xrightarrow{p} f(y_{t+1}| Y_t, \theta)$, and $\overline{W}_{t+1} \xrightarrow{p} F(y_{t+1}| Y_t, \theta)$.
Therefore, it follows that
\begin{equation*}
    \sum_{t=1}^n \log \overline{w}_t \xrightarrow{p} \sum_{t=1}^n \log f(y_t| y_1, \dots, y_{t-1}, \theta), 
\end{equation*}
is a consistent estimate of the conditional log-likelihood and can be used as an input in the calculation of the marginal likelihood by the method of \cite{Chib(95)}, as extended to M-H sampler output by \cite{ChibJeliazkov(01)}. 
\section{Illustrative numerical examples}
\label{sec:Illustrative examples}
This section illustrates our proposed estimation method using the simulated data. We generate $y_t$ $(t = 1, \dots, 1000)$ by setting
\begin{equation*}
    \phi = 0.97, \quad
    \mu = 0, \quad
    \sigma = 0.3.
\end{equation*} 
To avoid the case $y_t = 0$ which leads to $\log (y_t^2) = -\infty$, we introduce very small value $c$ and use $y_t^* = \log (y_t^2 + c)$. 
We set $c$ equal to $1.0 \times 10^{-7}$.
%
%
For $\beta$, we consider three cases $\beta = 0.3, 0.5$ and $0.7$ to investigate the effect of the approximation error.
The common random numbers are used to generate $y_t$'s for three cases. 
In these simulation studies, we specify the prior as
\begin{align*}
    &\mu \sim N(0, 3^2), \quad \frac{\phi + 1}{2} \sim Beta(1,1), 
    \\
    & \sigma^2 \sim IG\left( \frac{0.001}{2}, \quad 
    \frac{0.001}{2} \right), \quad 
     \beta  \sim N \left( 0, 1 \right).
\end{align*}
The prior on $(\phi+1)/2$ is set to ensure the stationarity of the latent volatility process. 
We iterated MCMC simulation 50,000 times after discarding initial 10,000 MCMC draws as burn-in period.
\begin{itemize}
    \item[(i)] Case $\beta = 0.3$. 
    The acceptance rates of the MH algorithms for $\alpha$ is 72.8\%. 
    The sample paths are given in Figure \ref{fig:sim0_3_samplepath}, and the MCMC chain mixes quite well. 
    \begin{figure}[H]
        \centering
        \includegraphics[width=0.7\linewidth]{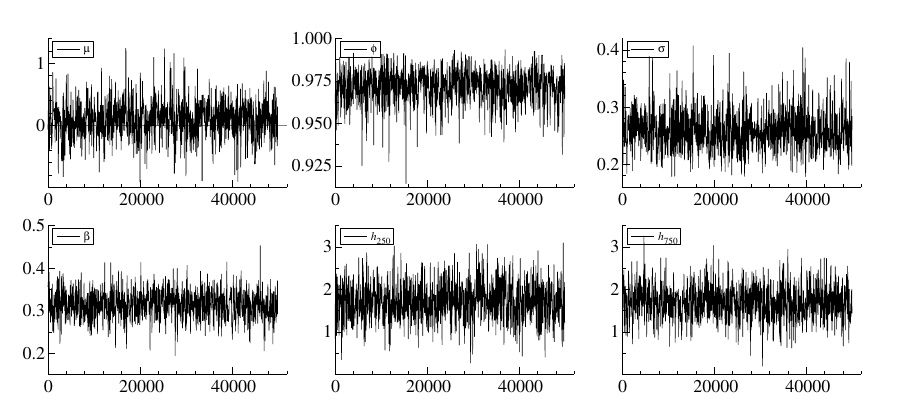}
        \caption{Sample paths for $\theta$, $h_{250}$, and $h_{750}$. $\beta = 0.3$.}
        \label{fig:sim0_3_samplepath}
    \end{figure}
    The sample autocorrelation functions are shown in Figure \ref{fig:sim0_3_ACF} and they decay very quickly.
    \begin{figure}[H]
    \centering
    \includegraphics[width=0.7\linewidth]{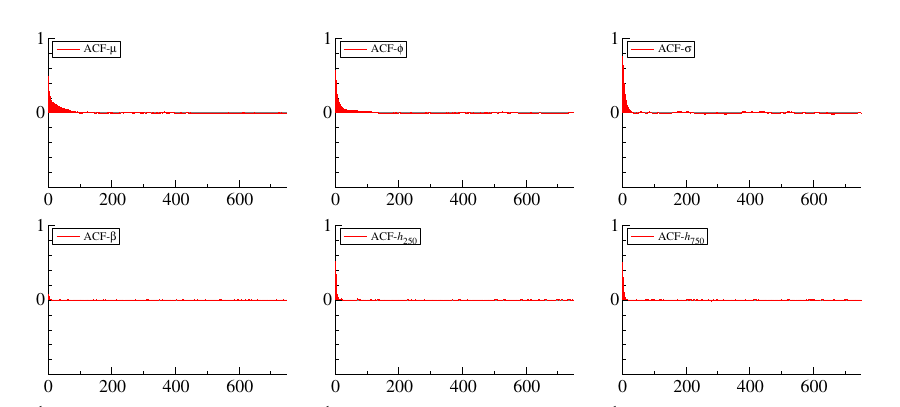}
    \caption{Sample autocorrelation functions for $\theta$, $h_{250}$, and $h_{750}$. $\beta = 0.3$.}
    \label{fig:sim0_3_ACF}
    \end{figure}
    \begin{table}[H]
    \small
      \centering
      \begin{tabular}{lrrrcr}
        \hline 
        Par & True & Mean & Std Dev & 95\% interval & IF \\
        \hline
        $\mu$ & 0 & 0.091 & 0.298 & (-0.514,  0.673) & 5 \\
        $\phi$ & 0.97 & 0.971 & 0.011 & ( 0.947,  0.988) & 5 \\
        $\sigma$ & 0.3 & 0.261 & 0.038 & ( 0.195,  0.344) & 10 \\
        $\beta$ & 0.3 & 0.316 & 0.033 & ( 0.251,  0.380) & 1 \\
        $h_{250}$ & 2.310 & 1.729 & 0.459 & ( 0.852,  2.651) & 4 \\
        $h_{750}$ & 2.077 & 1.675 & 0.416 & ( 0.888,  2.516) & 2 \\
        \hline
      \end{tabular}
      \caption{True values, posterior means, posterior standard deviations, 95\% credible intervals, and inefficiency factors. $\beta = 0.3$.}
      \label{table:sim0_3_MCMC}
      \normalsize
    \end{table}
    Table \ref{table:sim0_3_MCMC} shows the posterior mean, 95\% credible intervals and inefficiency factors (IF). 
    The estimated parameters are close to true values.
    IF is calculated by $1 + 2\sum_{s=1}^\infty \rho_s$ where $\rho_s$ is the sample autocorrelation at lag $s$.
    This is interpreted as the ratio of the numerical variance of the posterior mean from the chain to the variance of the posterior mean from hypothetical uncorrelated draws.
    They are overall small as expected, which means that the MCMC sampling is close to the uncorrelated sampling. 
    Note that those IF's for $h_{250}$ and $h_{750}$ are quite small, which suggests the use of mixture sampler for the MH algorithm for $h$ is highly efficient.
    Finally Figure \ref{fig:sim0_3_plot} shows true values, 95\% credible intervals and the posterior medians or volatilities. 
    The estimated smoothed values follow the true values values that are almost covered by 95\% intervals, indicating that MCMC estimations works well.

    \begin{figure}[H]
        \centering
        \includegraphics[width=0.8\linewidth]{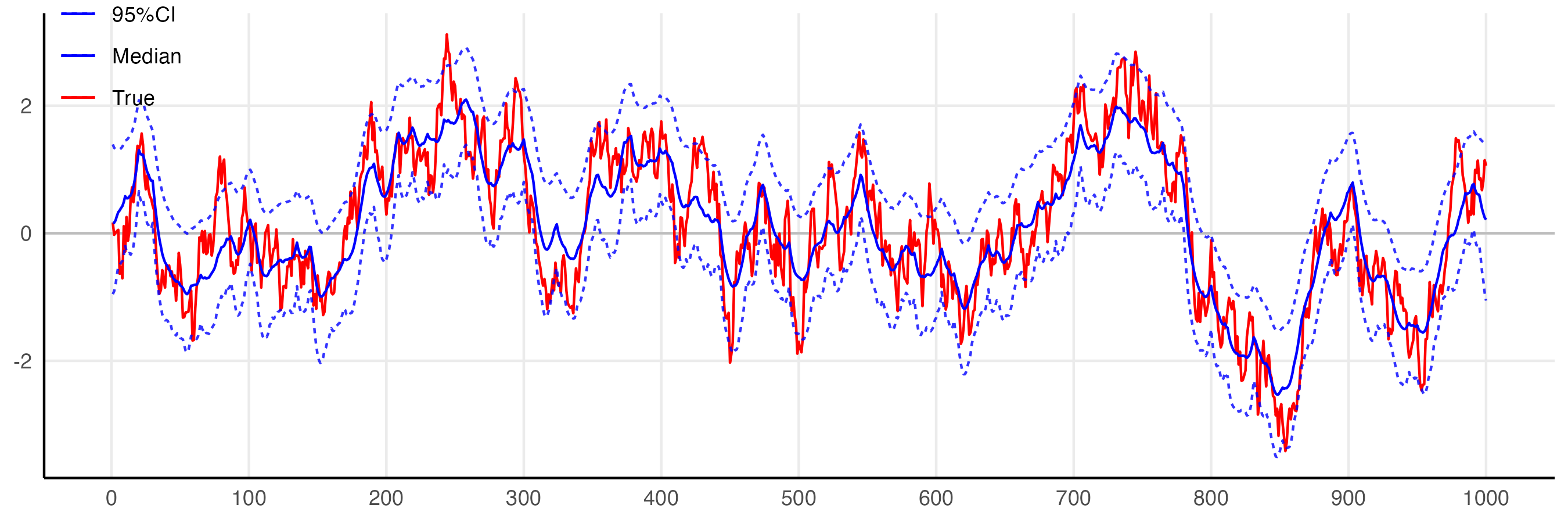}
        \caption{Log volatilities: True values, 95\% credible intervals and posterior median.}
        \label{fig:sim0_3_plot}
    \end{figure}    
    \item[(ii)] Case $\beta = 0.5$.
    The acceptance rates of the MH algorithms  for $\alpha$ is 73.2\%. 
    The plot of the sample paths and log volatilities are similar to those in (i) and hence omitted to save space. 
    Table \ref{table:sim0_5_MCMC} shows the posterior means, 95\% credible intervals and inefficiency factors. 
    The estimated parameters are close to true values, and inefficiency factors (IF) are overall small as in (i). 
    The IF's for $h_{250}$ and $h_{750}$ are sufficiently small, and indicates that the algorithm is still highly efficient.
    \begin{table}[H]
    \small
      \centering
      \begin{tabular}{lrrrcr}
        \hline 
        Par & True & Mean & Std Dev & 95\% interval & IF \\
        \hline
        $\mu$ & 0 & 0.104 & 0.319 & (-0.548,  0.734) & 31 \\
        $\phi$ & 0.97 & 0.971 & 0.011 & ( 0.948,  0.988) & 13 \\
        $\sigma$ & 0.3 & 0.262 & 0.036 & ( 0.198,  0.339) & 15 \\
        $\beta$ & 0.5 & 0.511 & 0.035 & ( 0.443,  0.579) & 2 \\
        $h_{250}$ & 2.310 & 1.797 & 0.461 & ( 0.903,  2.720) & 4 \\
        $h_{750}$ & 2.077 & 1.645 & 0.412 & ( 0.875,  2.493) & 4 \\
        \hline
      \end{tabular}
      \caption{True values, posterior means, posterior standard deviations, 95\% credible intervals, and inefficiency factors. $\beta = 0.5$.}
      \label{table:sim0_5_MCMC}
      \normalsize
    \end{table}

    \item[(iii)] Case $\beta = 0.7$.
    The acceptance rates of the MH algorithms  for $\alpha$ is 75.2\%. 
    The convergence seems to become slightly slower, but the chain mixes well. 
    The plot of the sample paths and log volatilities are similar to those in (i) and hence omitted to save space. 
    Table \ref{table:sim0_7_MCMC} shows the posterior means, 95\% credible intervals and inefficiency factors. 
    The estimated parameters are close to true values, and inefficiency factors (IF) are overall relatively small.
    The IF's for $h_{250}$ and $h_{750}$ are small, and indicates that the algorithm is still works well.
    \begin{table}[H]
    \small
      \centering
      \begin{tabular}{lrrrcr}
        \hline 
        Par & True & Mean & Std Dev & 95\% interval & IF \\
        \hline
        $\mu$ & 0 & 0.114 & 0.304 & (-0.510,  0.703) & 5 \\
        $\phi$ & 0.97 & 0.971 & 0.010 & ( 0.948,  0.988) & 6 \\
        $\sigma$ & 0.3 & 0.266 & 0.036 & ( 0.202,  0.342) & 9 \\
        $\beta$ & 0.7 & 0.704 & 0.037 & ( 0.633,  0.776) & 3 \\
        $h_{250}$ & 2.310 & 1.874 & 0.452 & ( 1.008,  2.779) & 2 \\
        $h_{750}$ & 2.077 & 1.661 & 0.407 & ( 0.898,  2.500) & 2 \\
        \hline
      \end{tabular}
      \caption{True values, posterior means, posterior standard deviations, 95\% credible intervals, and inefficiency factors. $\beta = 0.7$.}
      \label{table:sim0_7_MCMC}
      \normalsize
    \end{table}
\end{itemize}
These simulation results show that our proposed sampling method works well for those $\beta$'s found in the past empirical studies. 

\subsubsection*{Comparison of sampling efficiencies}
Next, we compare the sampling efficiency of our proposed method with the method in \cite{Chan(17)}, which was shown to be more efficient than the more complex methods of \cite{Mccausland(12)} and \cite{Andrieu(10)}. Chan's method needs to implement the accept-reject Metropolis-Hastings (ARMH) algorithm \citep{ChibGreenberg95} (the usual MH algorithm could be used, but then the rejection rate becomes much higher), using a proposal distribution based on a multivariate normal approximation of the full conditional distribution, obtained by a second-order Taylor expansion.

Table \ref{table:IFs} shows the IFs for selected values ($h_t$, $t=100,200,\ldots,1000$), the mean ($\overline{h}$) and the median ($h_{med}$) of $h_t$ for $\beta=0.3, 0.5$ and $0.7$. Among the three algorithms, the generalized mixture sampler (Algorithm 1, denoted by GMS) is the most efficient with IFs less than 10. These are substantially smaller than those of the CHN method. The IFs for Algorithm 2, the generalized mixture sampler with an additional Metropolis-Hastings step (denoted as GMH), are somewhat larger, but still smaller than those from Chan's method (denoted by CHN). We note that the IFs of GMH become larger as the absolute value of $\beta$ increases. However, as shown in the results of the simulation experiment above and in Appendix \ref{appendix:correction}, the resulting posterior estimates by GMS and GMH are almost the same. The takeaway from this experiment is the striking efficiency of the GMS method. 

In addition, we compare the computational times required in each experiment for the three algorithms. As shown in Table \ref{table:computation_time}, our proposed methods (GMS and GMH) are much faster than Chan's method (CHN). This is mainly because (i) GMS and GMH sample $h$ using the highly efficient and fast simulation smoother, whereas the CHN method is based on an expensive tailoring step that entails the inversion of a high dimensional covariance matrix in each MCMC iteration; (ii) the AR step of the ARMH algorithm degrades with many rejections when $h$ is high-dimensional and $\beta$ is large; and (iii) the search for the mode in the tailoring step by an iterative optimization methods consumes considerable time for longer time series (i.e., as the dimension of $h$ increases). Our proposed methods encounters none of these difficulties. 
\begin{table}[H]
  \small
  \centering
  \begin{tabular*}{\textwidth}{l@{\extracolsep{\fill}}*{9}{r}}
    \hline 
     & \multicolumn{3}{c}{$\beta=0.3$} 
     & \multicolumn{3}{c}{$\beta=0.5$}
     & \multicolumn{3}{c}{$\beta=0.7$} \\
     \cline{2-4}   \cline{5-7} \cline{8-10}  
     $h_t$ & GMS & GMH & CHN & GMS & GMH & CHN & GMS & GMH & CHN \\
     \hline
    $h_{100}$ & 6 & 11 & 32 & 2 & 26 & 18 & 2 & 94 & 87 \\
    $h_{200}$ & 4 & 32 & 50 & 5 & 21 & 40 & 6 & 85 & 89 \\
    $h_{300}$ & 2 & 12 & 22 & 7 & 31 & 43 & 3 & 116 & 73 \\
    $h_{400}$ & 4 & 23 & 67 & 3 & 41 & 63 & 4 & 82 & 65 \\
    $h_{500}$ & 8 & 10 & 43 & 6 & 32 & 60 & 6 & 88 & 209 \\
    $h_{600}$ & 1 & 17 & 41 & 2 & 35 & 48 & 2 & 38 & 40 \\
    $h_{700}$ & 5 & 18 & 50 & 5 & 26 & 15 & 4 & 51 & 64 \\
    $h_{800}$ & 4 & 12 & 38 & 6 & 71 & 14 & 3 & 56 & 103 \\
    $h_{900}$ & 3 & 11 & 27 & 7 & 48 & 30 & 4 & 25 & 26 \\
    $h_{1000}$ & 2 & 6 & 23 & 3 & 35 & 42 & 5 & 114 & 99 \\
    $\overline{h}$ & 8 & 28 & 182 & 9 & 68 & 105 & 9 & 135 & 165 \\
    $h_{med}$ & 7 & 15 & 74 & 7 & 25 & 35 & 4 & 62 & 70 \\
    \hline
\end{tabular*}
  \caption{Inefficiency factors of selected $h_t$'s with their mean ($\overline{h}$) and median ($h_{med}$) in our simulation studies. Generalized mixture sampler (GMS), GMS with MH step for approximation correction (GMH), and Chan's method (\cite{Chan(17)}) (CHN).}
  \label{table:IFs}
  \normalsize
\end{table}

\begin{table}[H]
  \small
  \centering
  \begin{tabular*}{\textwidth}{l@{\extracolsep{\fill}}*{9}{r}}
    \hline 
     & \multicolumn{3}{c}{$\beta=0.3$} 
     & \multicolumn{3}{c}{$\beta=0.5$}
     & \multicolumn{3}{c}{$\beta=0.7$} \\
     \cline{2-4}   \cline{5-7} \cline{8-10}  
        & GMS & GMH & CHN & GMS & GMH & CHN & GMS & GMH & CHN \\
     \hline
    Time & 1,792 & 1,910 & 133,884 & 1,701 & 1,935 & 134,127& 1,700 & 1,948 & 126,168\\
    \hline
  \end{tabular*}
  \caption{Computational time (seconds) for our simulation studies. We draw 50,000 MCMC samples after discarding initial 10,000 samples as burn-in period.}
  \label{table:computation_time}
  \normalsize
\end{table}
\section{Extension to SVM model with Leverage (SVML model)}
\label{sec:approx_SVM_leverage}
In this section, we consider the SVM model with leverage which we call SVML model. The leverage effect implies the decrease in the return at time $t$  followed by the increase in the  volatility at time $t+1$. Thus we incorporate the correlation $\rho$ between $y_t$ and $h_{t+1}$ and replace (\ref{eq:error_without_leverage}) by
\begin{align} 
        &\begin{pmatrix}
            \epsilon_t \\
            \eta_t
        \end{pmatrix}
        \overset{\text{i.i.d.}}{\sim} N(0, \Sigma), \quad
        \Sigma = 
        \begin{pmatrix}
            1 & \rho \sigma \\
            \rho \sigma & \sigma^2
        \end{pmatrix}.  \label{eq:error_with_leverage}
\end{align}
The negative correlation, $\rho<0$, indicates the existence of the leverage effect. 
%
%
Next, we construct the linear and Gaussian state space model that approximates the SVM model with leverage using the mixture of the normal densities given in (\ref{approx_noncentral_chi2}).
We first let $d_t = I(y_t \geq 0) - I(y_t < 0)$ where $I(A) = 1$ if $A$ is true and $I(A) = 0$ otherwise.
Noting that
\begin{align*}
    y_t =& d_t \exp(y_t^*/2), \quad 
    \epsilon_t = d_t \exp(\epsilon_t^*/2) - \beta \\
    &\eta_t|\epsilon_t \sim N(\rho \sigma \epsilon_t, \sigma^2(1–\rho^2)),
\end{align*}
we rewrite the conditional distribution as
\begin{equation*}
    \eta_t|\epsilon_t \sim N\left(\rho \sigma \{ d_t \exp(\epsilon_t^*/2) - \beta \}, \sigma^2(1–\rho^2)\right).
\end{equation*}
Let $s_t = (s_{1t}, s_{2t}) \in \{ (i,j) | i=1,...,K, j=0,...,J \}$ denote the component of the mixture of normal densities in (\ref{approx_noncentral_chi2}) at time $t$.
Given $s_t = (i,j)$, we have $\epsilon_t^* | s_t=(i,j) \sim N(\Tilde{m}_{i,j}, v_i^2)$.
Furthermore, we approximate $\exp(\epsilon_t^*/2)$ by $\exp(\Tilde{m}_{i,j}/2) \{ a_i + b_i(\epsilon_t^* - \Tilde{m}_{i,j}) \}$ with $a_i = \exp(v_i^2/8)$, $b_i = \frac{1}{2} \exp(v_i^2/8)$, as in Table \ref{table:approx_chi2}, which minimize the mean square norm
\begin{equation*}
    E[ \exp(\epsilon_t^*/2) - \exp(\Tilde{m}_{i,j}/2) \{ a_i + b_i(\epsilon_t^* - \Tilde{m}_{i,j}) \} ]^2.
\end{equation*}
Thus, the approximate conditional distribution is 
\begin{equation*}
    \eta_t|\epsilon_t \sim N\left(\rho \sigma [ d_t \exp(\Tilde{m}_{i,j}/2) \{ a_i + b_i(\epsilon_t^* - \Tilde{m}_{i,j}) \} - \beta ], \sigma^2(1–\rho^2)\right).
\end{equation*}
Given $s = (s_1, ..., s_n)$, we find that the SVM model can be approximated by the linear Gaussian state space form
\begin{align}
    y_t^* &= \Tilde{m}_{s_{1t}, s_{2t}} + h_t + (v_{s_{1t}}, 0) z_t, \label{approx_linaer_start} \\
    h_{t+1} &= \mu(1-\phi) + \rho \sigma \{ d_t a_{s_{1t}} \exp(\Tilde{m}_{s_{1t}, s_{2t}}) - \beta \} + \phi h_t \\
    &\quad + (d_t \rho \sigma b_{s_{1t}} v_{s_{1t}} \exp(\Tilde{m}_{s_{1t}, s_{2t}}/2), \sigma \sqrt{1-\rho^2}) z_t, \nonumber \\
    &t = 1,...,n-1, \quad h_1 \sim N\left( \mu, \frac{\sigma^2}{1-\phi^2} \right), \quad
    |\phi| < 1, \label{approx_linaer_end} \\
    &z_t = (z_{1t}, z_{2t})' \sim N(0, I_2), \nonumber
\end{align}
where $y_t^* = \log(y_t^2)$ and $d_t = I(y_t \geq 0) - I(y_t < 0)$. 
In the following subsections, $\Tilde{m}_{s_{1t}, s_{2t}}$ and $\Tilde{p}_{s_{1t}, s_{2t}}$ are abbreviated as $\Tilde{m}_{s_t}$ and $\Tilde{p}_{s_t}$, and we write $v_{s_t}^2, a_{s_t}$, and $b_{s_t}$ instead of $v_{s_{1t}}^2, a_{s_{1t}}$, and $b_{s_{1t}}$, respectively. The MCMC algorithm and the particle filter are detailed in Appendix \ref{appendix:mcmc and particle filter leverage}.
\bigskip

\noindent
\subsection*{Extension to the multivariate SVM model}
Furthermore, we will give a couple of examples to illustrate how to extend our proposed SVM model to the multivariate models. As a first example, consider the factor multivariate stochastic volatility (MSV) model proposed by  \cite{ChibNardariShephard(06)} (see \cite{IshiharaOmori(17)} for the model with leverage). Let $\bm{y}_t=(y_{1t},\ldots,y_{pt})'$ and $\bm{f}_t=(f_{1t},\ldots,f_{qt})'$ denote the dependent and factor variables ($q<p$) respectively. The basic factor MSV model is given by
\begin{eqnarray*}
    \bm{y}_{t} &=& \mathbf{B}\bm{f}_t + \mathbf{V}_{1t}^{1/2}\bm{\epsilon}_{1t}, \quad \bm{\epsilon}_{1t}\sim N(\bm{0}, \mathbf{I}_p),\\
    \bm{f}_{t} &=& \mathbf{V}_{2t}^{1/2}\bm{\epsilon}_{2t}, \quad \bm{\epsilon}_{2t}\sim N(\bm{0}, \mathbf{I}_q),\\
    \bm{h}_{t+1} &=& \bm{\mu}+\mathbf{\Phi}(\bm{h}_{t-1}-\bm{\mu})+\bm{\eta}_t, \quad \bm{\eta}_t\sim N(\bm{0}, \mathbf{\Sigma}), \\
    && \mathbf{V}_{1t}=\mbox{diag}(\exp(h_{1t}), \ldots, \exp(h_{pt})), 
    \quad \mathbf{V}_{2t}=\mbox{diag}(\exp(h_{p+1,t}), \ldots, \exp(h_{p+q,t})), \\
    && \mathbf{\Phi} = \mbox{diag}(\phi_1, \ldots, \phi_p), \quad \mathbf{\Sigma} = \mbox{diag}(\sigma_1^2,\ldots,\sigma_p^2,\sigma_{p+1}^2,\ldots\sigma_{p+q}^2),
\end{eqnarray*}
where $\mathbf{B}$ is a $p\times q$ factor loading matrix subject to constraints ($b_{ij}=0$ for $j>i$ and $b_{ii}=1$ for $i \leq q$). 
To incorporate the SVM in the MSV model, we replace the factor equation by
\begin{eqnarray*}
    \bm{f}_{t} &=& \mathbf{V}_{2t}^{1/2}\bm{\beta}+\mathbf{V}_{2t}^{1/2}\bm{\epsilon}_{2t},
\end{eqnarray*}
where $\bm{\beta}=(\beta_1,\ldots,\beta_q)'$ is a $q\times 1$ vector of weights.\\
\indent
As a second example,  we may consider the MSV model with the SVM that is common within a group. Let $\bm{y}_t=(y_{1t},\ldots,y_{pt})'$ and consider 
\begin{eqnarray*}  \bm{y}_{t}&=&\exp(\tilde{h}_t/2)\bm{\beta}+\exp(\tilde{h}_t/2)\bm{\epsilon}_t, \quad \bm{\epsilon}_t\sim N(\bm{0},\mathbf{\Sigma}),\\
\tilde{h}_{t+1}&=&\phi \tilde{h}_t + \eta_t, \quad \eta_t\sim N(0,\sigma^2),\quad\mathbf{\Sigma} = \mbox{diag}(\sigma_1^2,\ldots,\sigma_p^2),
\end{eqnarray*}
where $\bm{\beta}=(\beta_1,\ldots,\beta_p)'$. To linearize the measurement equation with respect to $\tilde{h}_t$, we define $\bm{y}^*=(y_{1t}^*,\ldots,y_{pt}^*)'$ and $\bm{\epsilon}_t^*=(\epsilon_{1t}^*,\ldots, \epsilon_{pt}^*)'$ where $y_{it}^*=\log y_{it}^2$ and $\epsilon_{it}^*=\log (\beta_i/\sigma_i+\epsilon_{it}/\sigma_i)^2$. Thus the transformed measurement equation is
\begin{eqnarray*}  
\bm{y}_t^*=\bm{\mu}^*+\tilde{h}_t\bm{1}_p+\bm{\epsilon}_t^*,\quad \bm{\mu}^*=(\log \sigma_1^2,\ldots,\log\sigma_p^2)', \quad \bm{1}_p=(1,\ldots,1)',
\end{eqnarray*}
Noting that $\epsilon_{it}^*\sim \log \chi_1^2(\beta_i^2/\sigma_i^2)$, we can approximate the noncentral chisquare distribution by the mixture of normal distributions as we have described.
\bigskip

\noindent
{\it Remark}. 
Some previous studies considered alternative mean specifications of $y_t$ using $\exp(h_t)$ or $h_t$, as in \cite{EngelLilienRobins(87)} for ARCH-M model. We note that our proposed sampler could be utilized to generate $h_t's$ efficiently, but details are left for our future work. 

\section{Empirical studies of excess holding yield data} \label{Empirical studies}
\subsection{Data}
\begin{figure}[H]
    \centering
    \includegraphics[width=0.6\linewidth]{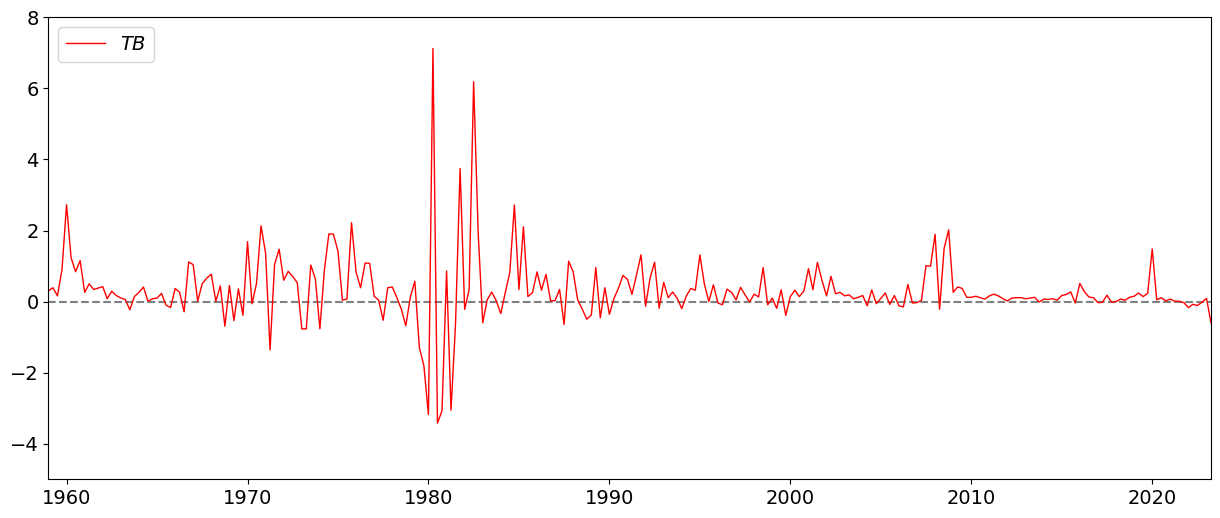}
    \caption{Time series plot. TB data: 1958Q4--2023Q1.}
    \label{fig:TB_raw}
\end{figure}

\begin{figure}[H]
    \centering
    \includegraphics[width=0.6\linewidth]{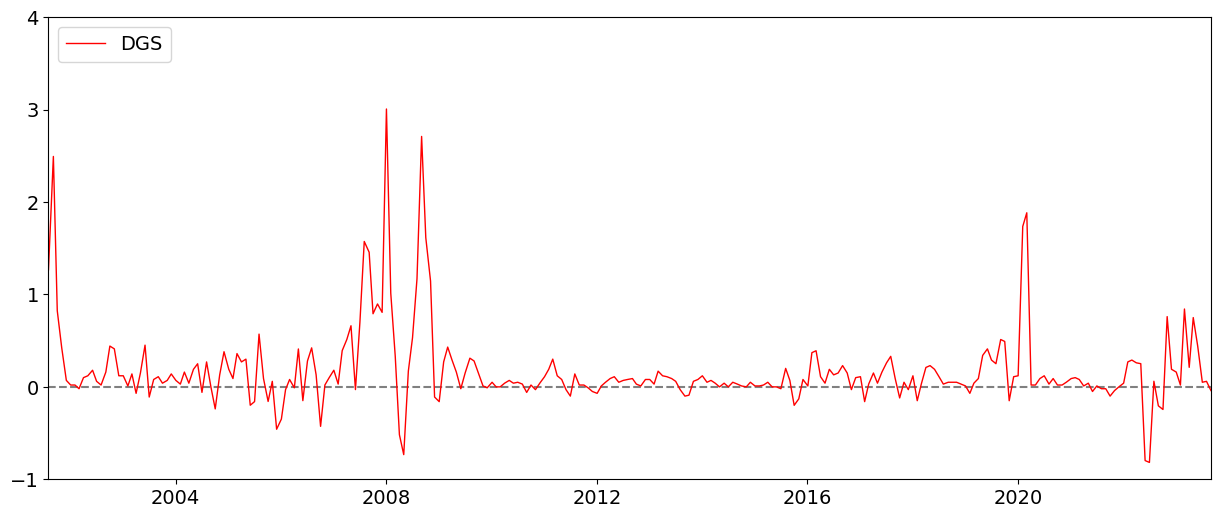}
    \caption{Time series plot. DGS data: 2001/8--2023/9.}
    \label{fig:DGS_raw}
\end{figure}

\begin{figure}[H]
    \centering
    \includegraphics[width=0.6\linewidth]{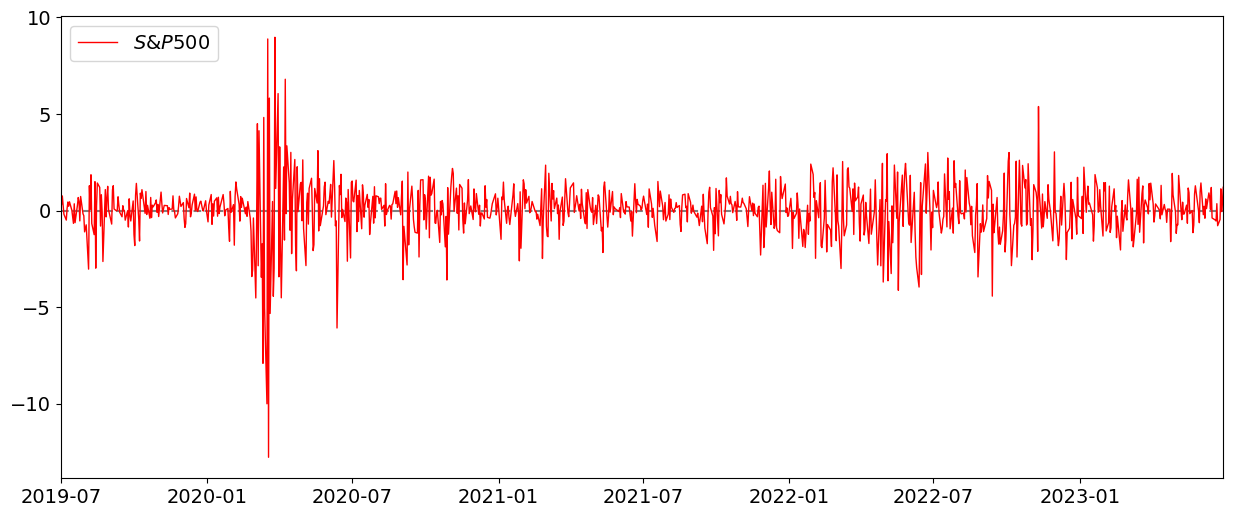}
    \caption{Time series plot. S\&P500 data: 2019/7/1--2023/6/30.}
    \label{fig:SP500_raw}
\end{figure}

This section applies the SVM model with leverage and several alternative models to three excess holding yields data. The descriptions of the data (labeled as TB, DGS and S\&P500) are given below\footnote{The data are obtained from the website of Federal Reserve Bank of St. Louis.}.
\begin{enumerate}
    \item[(1)] TB: the excess holding yield using 3 and 6 months treasury bills with 258 observations from the forth quarter of 1958 to the first of 2023. It is defined as
    \begin{eqnarray*}
         y_t = \left\{\frac{\left(1+\frac{R_t}{100}\right)^2}{1+\frac{r_{t+1}}{100}} -\left(1+\frac{r_t}{100}\right)\right\}\times 100, 
    \end{eqnarray*}
    at annual rate where $R_t$ and $r_t$ are secondary market rates of the 6-month and 3-month Treasury bill (discount basis, percent, daily, not seasonally adjusted), measured at the beginning of the quarter.
  
    \item[(2)] DGS: the excess holding yield using 1 and 3 month market yields on U.S. treasury securities with 266 observations from August 2001 to September 2023. The excess holding yield, $y_t$ is defined as
        \begin{eqnarray*}
         y_t = \left\{\frac{\left(1+\frac{R_t}{100}\right)^3}{\left(1+\frac{r_{t+1}}{100}\right)\left(1+\frac{r_{t+2}}{100}\right)} -\left(1+\frac{r_t}{100}\right)\right\}\times 100, 
    \end{eqnarray*}
    at annual rate where $R_t$ and $r_t$ are market yields of 3 and 1 month on US Treasury securities at constant maturity of 3 months (quoted on an investment basis, percent, daily, not seasonally adjusted), measured at the beginning of the month.
    
    \item[(3)] S\&P500: the excess return using S\&P500 index daily return and federal funds rate with 1008 observations from July 1st of 2019 to June 30 of 2023. It is defined as $y_t = R_t-r_t$ at daily rate where $R_t$ and $r_t$ are the daily log return of S\&P500 (in percent), and federal funds effective rate (percent, daily, not seasonally adjusted) divided by 360.
\end{enumerate}
The time series plots of three datasets are shown in Figures \ref{fig:TB_raw}, \ref{fig:DGS_raw} and \ref{fig:SP500_raw}. Volatility clustering phenomena is observed for all three series, suggesting that the stochastic volatility models are appropriate to describe these excess holding yield data.


\subsection{Estimation results for SVM model}
\label{sec:Estimation results for SVM model}
Using the same prior distributions as in illustrative examples in Section \ref{sec:Illustrative examples}, the proposed SVM models with leverage are fitted to TB, DGS and S\&P500 data.  
We iterated MCMC simulation 50,000 times after discarding initial 10,000 MCMC draws as burn-in period using Algorithm 3 in Appendix \ref{appendix:mcmc and particle filter leverage}. The acceptance rates of the MH algorithms for $\alpha$ and $(\alpha, h)$ are 72.0\% and 20.8\% with TB data, 69.5\% and 12.8\% with DGS data, and 77.3\% and 58.9\% with S\&P500 data, respectively.

\begin{figure}[H]
    \centering
    \includegraphics[width=0.6\linewidth]{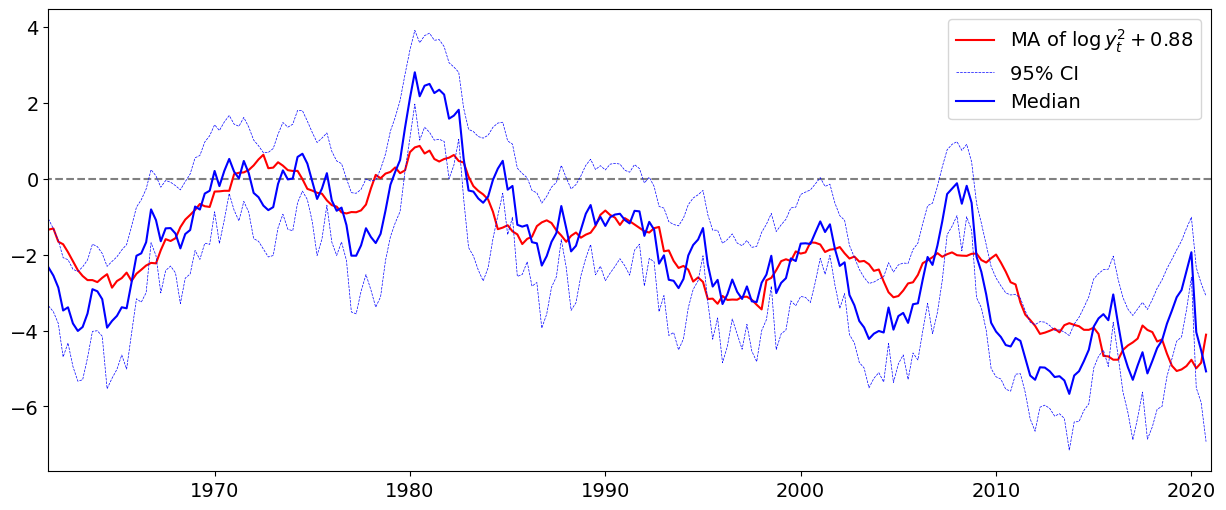}
    \caption{Log volatilities: Moving average of $\log (y_t^2) - E(\log \chi_1^2(\beta^2))$, 95\% credible intervals and posterior median. TB data.}
    \label{fig:TB_plot}
\end{figure}

\begin{figure}[H]
    \centering
    \includegraphics[width=0.6\linewidth]{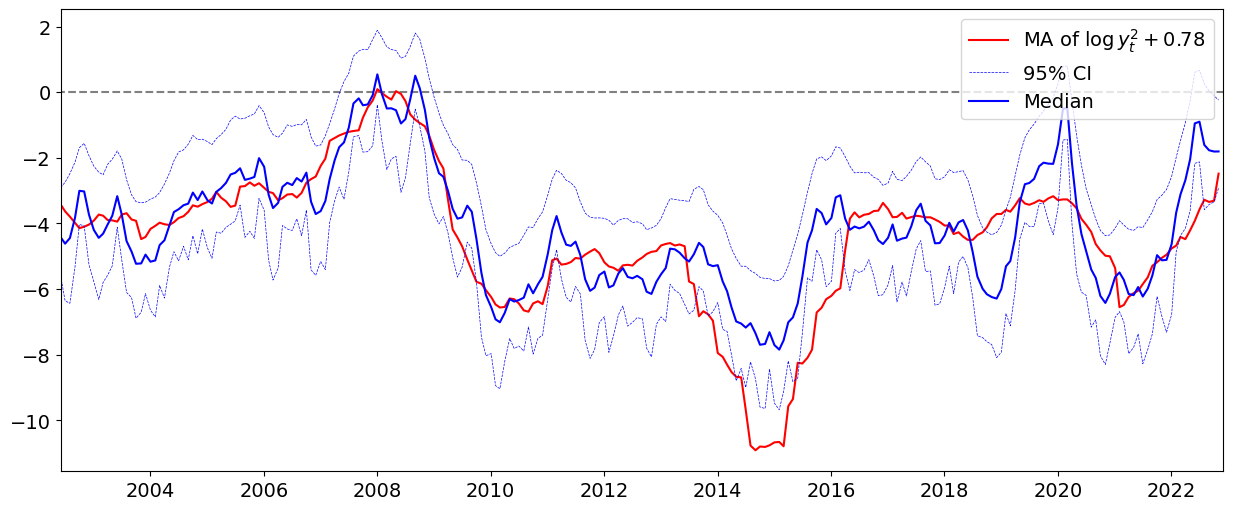}
    \caption{Log volatilities: Moving average of $\log (y_t^2) - E(\log \chi_1^2(\beta^2))$, 95\% credible intervals and posterior median. DGS data.}
    \label{fig:DGS_plot}
\end{figure}

\begin{figure}[H]
    \centering
    \includegraphics[width=0.6\linewidth]{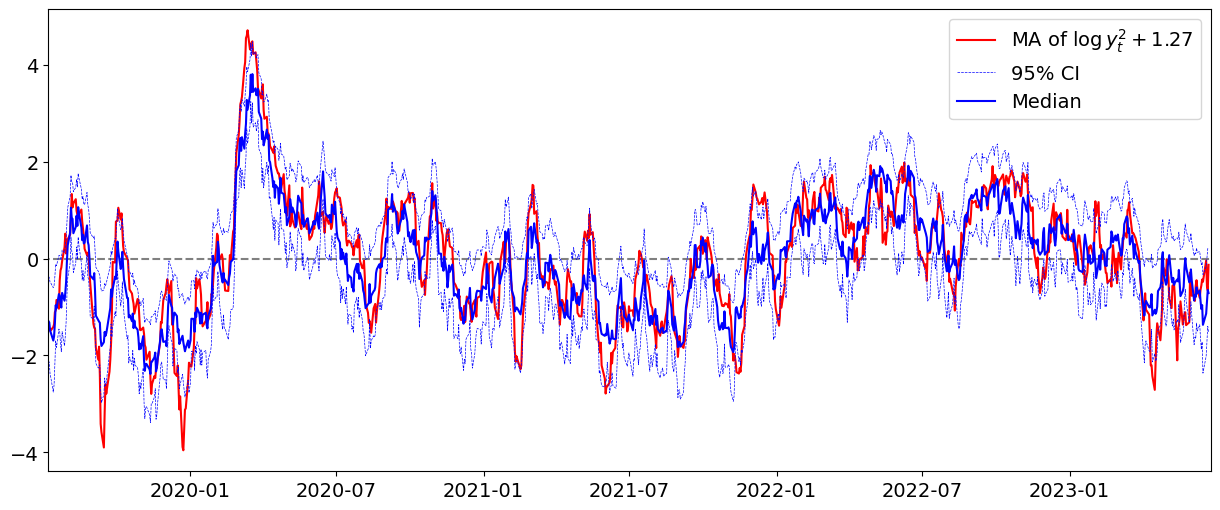}
    \caption{Log volatilities: Moving average of $\log (y_t^2) - E(\log \chi_1^2(\beta^2))$, 95\% credible intervals and posterior median. S\&P500 data.}
    \label{fig:SP500_plot}
\end{figure}

\begin{table}[H]
    \small
    \centering
    \begin{tabular}{lrrcrr}
    \hline 
    Par & Mean & Std Dev & 95\% interval & IF & Pr(+) \\
    \hline
    $\mu$ & -1.817 & 0.570 & (-2.941, -0.698) & 57 & 0.000 \\
     & -3.665 & 0.635 & (-4.763, -2.217) & 73 & 0.000 \\
     & -0.039 & 0.215 & (-0.474,  0.382) & 10 & 0.425 \\
    $\phi$ & 0.920 & 0.024 & ( 0.865,  0.960) & 22 & 1.000 \\
     & 0.903 & 0.038 & ( 0.813,  0.962) & 93 & 1.000 \\
     & 0.950 & 0.011 & ( 0.927,  0.969) & 14 & 1.000 \\
    $\sigma$ & 0.685 & 0.097 & ( 0.512,  0.894) & 18 & 1.000 \\
     & 0.911 & 0.120 & ( 0.698,  1.179) & 46 & 1.000 \\
     & 0.327 & 0.037 & ( 0.260,  0.404) & 21 & 1.000 \\
    $\rho$ & -0.546 & 0.139 & (-0.784, -0.252) & 87 & 0.000 \\
     & 0.051 & 0.129 & (-0.207,  0.305) & 37 & 0.659 \\
     & -0.698 & 0.062 & (-0.805, -0.563) & 11 & 0.000 \\
    $\beta$ & 0.649 & 0.073 & ( 0.507,  0.793) & 29 & 1.000 \\
     & 0.734 & 0.076 & ( 0.587,  0.884) & 47 & 1.000 \\
     & 0.060 & 0.032 & (-0.003,  0.124) & 5 & 0.969 \\
    $h_{100}$ & -0.648 & 0.977 & (-2.744,  1.108) & 18 & 0.259 \\
     & -5.490 & 1.018 & (-7.498, -3.576) & 49 & 0.000 \\
     & -2.252 & 0.511 & (-3.251, -1.248) & 4 & 0.000 \\
    \hline
  \end{tabular}
  \caption{Posterior mean, standard deviation, 95\% credible interval, inefficient factor and the posterior probability that the parameter is positive. TB (top row), DGS (middle row), and S\&P500 (bottom row).}
  \label{table:TB_DGS_SP500_MCMC}
    \normalsize
\end{table}
Table \ref{table:TB_DGS_SP500_MCMC} shows posterior means, standard deviations, 95\% credible intervals, inefficiency factors for parameters (IF), and the posterior probability that the parameter is positive for three datasets. 
Further, the IF's for log volatilities, $h_t$'s are found to be less than 80, which implies that our mixture sampler is highly efficient as shown in Section \ref{sec:Illustrative examples}. 
The coefficient $\beta$ is estimated to be greater than 0.6 for TB and DGS data, albeit close to zero for S\&P500 data. In all cases, we find strong evidence of the positive risk premium since the posterior probability that $\beta$ is positive is almost one for TB and DGS data and 0.97 for S\&P500 data.
The autoregressive parameter $\phi$ for the log volatility process is estimated to be more than 0.9, suggesting the high persistence in the volatility as found in the various empirical studies in the previous literature.
The correlation parameter $\rho$ is estimated to be negative for TB data and S\&P500 data, implying a strong evidence of the leverage effect since the posterior probability that $\rho$ is negative is almost one, $Pr(\rho < 0|y) \approx 1.000$. On the other hand, there is no evidence that $\rho$ is negative for DGS data.
\noindent
Finally, Figures \ref{fig:TB_plot}, \ref{fig:DGS_plot} and \ref{fig:SP500_plot} show 95\% credible intervals and posterior medians for $h$ in the case of TB, DGS and S\&P500 data. 
Noting that
\begin{equation*}
    \log (y_t^2) = h_t + \log \chi_1^2(\beta^2),
\end{equation*}
we further plotted moving average $\sum_{s=-10}^{10} z_{t+s}/21$ for reference where $z_t = \log (y_t^2) - E(\log \chi_1^2(\beta^2))$ is evaluated at the posterior means of $\beta$. The expected values of $\log \chi_1^2(\beta^2)$ are computed numerically as $-0.88, -0.78$, and $-1.27$ for TB, DGS, and S\&P500 using Monte Carlo integration. The traceplot of the estimated log volatilities is similar to that of the moving average series taking account of 95\% credible intervals. The large volatilities around years 1980, 2008 and 2020/3-2020/4 are well captured by the proposed model as shown in Figures \ref{fig:TB_plot},  \ref{fig:DGS_plot} and  \ref{fig:SP500_plot} respectively.

\subsection{Model comparison}
\label{sec:alternative-models}
In this section we use Bayesian marginal likelihoods to conduct a comparison of different stochastic volatility models. We calculate the marginal likelihood using the method of \cite{Chib(95)}. Suppressing the model index, this method is based on an identity introduced in that paper: 
\begin{equation*}
    \log m(y) = \log f(y|\theta) + \log \pi(\theta) - \log \pi(\theta|y), 
\end{equation*}
where the first term on the right side is the log of the likelihood. the second term is the prior, and the third is the posterior density. We evaluate each of these terms with the posterior mean of $\theta$.
For each model, we calculate the first term using the particle filter method given in Section \ref{partile filter}, setting $I = 80,000$. To compute the posterior density ordinate, we apply \cite{ChibJeliazkov(01)} to the MCMC draws from Algorithm 4 in Appendix \ref{appendix:posterior ordinate}.

The SVML model had the highest log marginal likelihood for TB and S\&P500 data, while the SVM model had the highest for DGS data. This implies our proposed model best describes the risk premium and the time varying volatility among competing models including the standard SV and SVL models.
These results are also consistent with high posterior probabilities of $Pr(\beta > 0|y) $ for TB, DGS and S\&P500 data, and of $Pr(\rho< 0|y)$ for TB and S\&P500 data as given in Table \ref{table:TB_DGS_SP500_MCMC} of Section \ref{sec:Estimation results for SVM model}.
\begin{table}[H]
    \small
  \centering
  \begin{tabular}{lrrr}
    \hline 
    Model & \multicolumn{1}{c}{TB} &\multicolumn{1}{c}{DGS} & \multicolumn{1}{c}{S\&P500} \\
    \hline
    SVM & -186.132(0.019) & \textbf{68.422(0.016)} & -1533.027(0.075) \\
    SVML & \textbf{-181.449(0.020)} & 66.695(0.015) & \textbf{-1508.296(0.059)} \\
    SV & -231.392(0.020) & 16.954(0.020) & -1536.776(0.060) \\
    SVL & -223.653(0.032) & 16.150(0.021) & -1509.656(0.039) \\
    \hline
  \end{tabular}
  \caption{Log marginal likelihood estimation and standard error (in parentheses). TB, DGS, and S\&P500 data. SVML 
  and SVL models include the leverage effect $\rho$. The bold font indicates the largest marginal likelihood.}
  \label{table:log_marginal_app}
  \normalsize
\end{table}
\section{Conclusion}
In this paper, we have successfully extended the mixture sampler for the SV model to the SVM model of which the mean equation is described by the standard deviation of the error term as an independent variable.
Our main point is the approximation of the distribution of $\log \chi_1^2(\beta^2)$ by mixture of normal distributions which is dependent on the parameter $\beta$. This approximation facilitates efficient sampling, leveraging well-established methods for the linear Gaussian state-space model. It is shown in simulation studies that our proposed method is implemented easily and works fast and efficiently.
In the empirical studies of the excess holding yield data and S\&P500 data, we conducted the model comparison among the SV and SVM models with and without leverage, and found that our proposed SVM models outperform other models in terms of the marginal likelihood. 
It shows that there exists the positive risk premiums and time-varying volatilities for all data, while the leverage effects are found to exist for TB and S\&P500 data. 
\section*{Acknowledgement}
The authors thank the editor and anonymous referees for their helpful comments. 
This work is partially supported by  JSPS KAKENHI [Grant number: 24H00142]. The
computational results are obtained using Rcpp and Ox (see \cite{Doornik(07)}).
\small
\bibliography{ref_ASVM_paper}	
\normalsize
%
\appendix
\section*{Appendix}
\section{MH step to correct the approximation error}
\label{appendix:correction}
\noindent

The posterior density is given by
\begin{align*}
    \pi(h, \theta|y) &\propto f(y, h|\theta) \pi(\theta) \\
    &\propto (1+\phi)^{a-\frac{1}{2}} (1-\phi)^{b-\frac{1}{2}}  (\sigma^2)^{-\left( \frac{n_1}{2}+1 \right)} \exp \left\{ -\frac{1}{2\sigma^2} \{ s_0 + (1-\phi^2)(h_1-\mu)^2 \} \right\} \\
    &\quad \times \exp \left\{ -\frac{1}{2} \sum_{t=1}^n [ h_t + \{ y_t -\beta\exp(h_t/2) \}^2 \exp(-h_t) ] \right\} \\
    &\quad \times \exp \left\{ -\frac{1}{2\sigma^2} \sum_{t=1}^{n-1} [ h_{t+1} - \mu(1-\phi) -\phi h_t ]^2 \right\} \\
    &\quad \times \exp \left\{ -\frac{(\mu-\mu_0)^2}{2\sigma_0^2} \right\} \exp \left\{ -\frac{(\beta-b_0)^2}{2B_0} \right\},
\end{align*}
where $n_1 = n_0 + n$.
To correct the approximation error in Step 3 of Algorithm 1, we implement the additional MH step (Step 4) as in the following Algorithm 2. \\ 
\bigskip

\noindent
\textbf{Algorithm 2 (GMS+MH algorithm, GMH).}
Let us denote $\theta = (\alpha,\beta)$ where $\alpha = (\mu,\phi,\sigma^2)$.
The Markov chain Monte Carlo simulation is implemented in four blocks:

\begin{itemize}
    \item[1.] Initialize $h$ and $\theta=(\alpha,\beta)$.
    \item[2.] Generate $\beta|\alpha,h, y \sim \pi(\beta|\alpha,h, y)$ as in Algorithm 1.
    \item[3.] Generate $(\alpha,h)|\beta, y \sim \pi(\alpha,h|\beta, y)$ as in Algorithm 1.
    \item[4.] Conduct MH algorithm to correct the approximation error.
    \item[5.] Go to step 2.
\end{itemize}
\subsubsection*{Step 4. Generation of $(\alpha,h)|\beta,y$}
%
Since the mixture sampler is based on the approximation, we can correct the approximation error after the MCMC simulation as in \cite{KimShephardChib(98)} and  \cite{OmoriChibShephardNakajima(07)}) Instead, we use the data augmentation method to correct it within the MCMC simulation by MH algorithm with the pseudo target density. We note that the similar approach has been considered for the SV model without leverage (\cite{DelPrimiceri(15)}) and with leverage (\cite{TakahashiOmoriWatanabe(23)}). 
Define the pseudo target density
\begin{align*}
    \Tilde{\pi}(\alpha,h,s|\beta,y) &= \pi(\alpha, h|\beta,y) \times q(s|h,\alpha,\beta, y^*), \\
    &q(s|h,\alpha,\beta, y^*) = \prod_{t=1}^n \frac{\Tilde{p}_{s_t} g(y_t^*| h_t, \alpha,\beta, s_t)}{\sum_{i=1}^{10}\sum_{j=0}^2 \Tilde{p}_{i,j} g(y_t^*| h_t, \alpha,\beta, s_t=(i,j))}, 
\end{align*}

\noindent
Note that the marginal density $\pi(\alpha,h|\beta,y)$ is our target density, $\pi(\alpha,h|\beta,y) = \sum_s \Tilde{\pi}(\alpha,h,s|\beta,y)$. We generate sample $(\alpha, h, s)$ from the pseudo target density as follows. Using Step 3 of Algorithm 1, we have a sample from $\pi^*(h|\alpha, s,\beta, y^*)\pi^*(\alpha|s,\beta,y)$ and let us denote it as $(\alpha^\dagger,h^\dag)$, and let
    \begin{align*}
        f(y_t|h_t, \alpha, \beta) = f_N(y_t| \beta \exp(h_t/2), \exp(h_t)), 
        \quad t=1,\ldots,n.
    \end{align*}
Given the current value $(\alpha,h)$, accept the candidate $(\alpha^\dagger,h^\dag)$ with probability
    \begin{align*}
        \min &\left\{ 1, \frac{\Tilde{\pi}(\alpha^\dag, h^\dag| s, \beta, y)\pi^*(h|\alpha, s, \beta,y ^*)\pi^*(\alpha|s,\beta,y)}{\Tilde{\pi}(\alpha, h|s, \beta,y)\pi^*(h^\dag|\alpha^{\dagger}, s, \beta, y^*)\pi^*(\alpha^{\dagger}|s,\beta,y)} \right\} \\
        &= \min \left\{ 1, \frac{\pi(\alpha^\dag, h^\dag| \beta,y) q(s|h^\dag,\alpha^\dag,\beta, y^*) \pi^*(h|\alpha, s,\beta, y^*)\pi^*(\alpha|s,\beta,y)}{\pi(\alpha,h|\beta,y) q(s|h,\alpha,\beta, y^*) \pi^*(h^\dag|\alpha^{\dagger}, s, \beta, y^*)\pi^*(\alpha^{\dagger}|s,\beta,y)} \right\} \\
        &= \min \left\{ 1, \frac{q(s|h^\dag,\alpha^\dag,\beta, y^*) \prod_{t=1}^n f(y_t| h_t^\dag, \alpha^\dag,\beta) g(y_t^*| h_t, \alpha,\beta, s_t) }{q(s|h,\alpha,\beta, y^*)\prod_{t=1}^n f(y_t| h_t, \alpha,\beta) g(y_t^*| h_t^\dag, \alpha^\dag,\beta, s_t) } \right\} \\
        &= \min \left\{ 1, \frac{\prod_{t=1}^n f(y_t| h_t^\dag, \alpha^\dag,\beta) \sum_{i=1}^{10}\sum_{j=0}^2 \Tilde{p}_{i,j} g(y_t^*| h_t, \alpha,\beta, s_t=(i,j)) }{\prod_{t=1}^n f(y_t| h_t, \alpha,\beta) \sum_{i=1}^{10}\sum_{j=0}^2 \Tilde{p}_{i,j} g(y_t^*| h_t^\dag, \alpha^\dag,\beta, s_t=(i,j)) } \right\}.
       \end{align*}
    \vspace{2mm}

    \noindent
    Below we compare estimates using Algorithms 1 and 2 in illustrative examples. The results are quite close to each other, implying that the approximation is highly accurate.
\begin{table}[H]
\footnotesize
  \centering
  \begin{tabular*}{\textwidth}{lr@{\extracolsep{\fill}}*{8}{r}}
    \hline 
     &  & \multicolumn{4}{c}{Algorithm 1 (GMS)} &  \multicolumn{4}{c}{Algorithm 2 (GMH)}  \\
     \cline{3-6} \cline{7-10}
    Par & True & Mean & Std Dev & 95\% interval & IF & Mean & Std Dev & 95\% interval & IF \\
    \hline
    $\mu$ & 0 & 0.091 & 0.298 & (-0.514,  0.673) & 5 & 0.085 & 0.316 & (-0.570,  0.742) & 31 \\
    $\phi$ & 0.97 & 0.971 & 0.011 & ( 0.947,  0.988) & 5 & 0.972 & 0.011 & ( 0.948,  0.989) & 24 \\
    $\sigma$ & 0.3 & 0.261 & 0.038 & ( 0.195,  0.344) & 10 & 0.259 & 0.037 & ( 0.195,  0.338) & 21 \\
    $\beta$ & 0.3 & 0.316 & 0.033 & ( 0.251,  0.380) & 1 & 0.327 & 0.033 & ( 0.263,  0.392) & 4 \\
    \hline
  \end{tabular*}
  \caption{True values, posterior means, posterior standard deviations, 95\% credible intervals, and inefficiency factors. $\beta = 0.3$.}
  \label{table:alg12_sim0_3_MCMC}
  \normalsize
\end{table}

\begin{table}[H]
\small
  \centering
  \begin{tabular*}{\textwidth}{lr@{\extracolsep{\fill}}*{8}{r}}
    \hline 
     &  & \multicolumn{4}{c}{Algorithm 1 (GMS)} &  \multicolumn{4}{c}{Algorithm 2 (GMH)}  \\
     \cline{3-6} \cline{7-10}   
     Par & True & Mean & Std Dev & 95\% interval & IF & Mean & Std Dev & 95\% interval & IF \\
    \hline
    $\mu$ & 0 & 0.104 & 0.319 & (-0.548,  0.734) & 31 & 0.084 & 0.339 & (-0.660,  0.779) & 80 \\
    $\phi$ & 0.97 & 0.971 & 0.011 & ( 0.948,  0.988) & 13 & 0.972 & 0.010 & ( 0.949,  0.989) & 61 \\
    $\sigma$ & 0.3 & 0.262 & 0.036 & ( 0.198,  0.339) & 15 & 0.258 & 0.035 & ( 0.198,  0.336) & 60 \\
    $\beta$ & 0.5 & 0.511 & 0.035 & ( 0.443,  0.579) & 2 & 0.530 & 0.034 & ( 0.463,  0.597) & 12 \\
    \hline
  \end{tabular*}
  \caption{True values, posterior means, posterior standard deviations, 95\% credible intervals, and inefficiency factors. $\beta = 0.5$.}
  \label{table:alg12_sim0_5_MCMC}
  \normalsize
\end{table}

\begin{table}[H]
\small
  \centering
  \begin{tabular*}{\textwidth}{lr@{\extracolsep{\fill}}*{8}{r}}    \hline 
     &  & \multicolumn{4}{c}{Algorithm 1 (GMS)} &  \multicolumn{4}{c}{Algorithm 2 (GMH)}  \\
     \cline{3-6} \cline{7-10}
    Par & True & Mean & Std Dev & 95\% interval & IF & Mean & Std Dev & 95\% interval & IF \\
    \hline
    $\mu$ & 0 & 0.114 & 0.304 & (-0.510,  0.703) & 5 & 0.103 & 0.319 & (-0.514,  0.744) & 90 \\
    $\phi$ & 0.97 & 0.971 & 0.010 & ( 0.948,  0.988) & 6 & 0.971 & 0.011 & ( 0.947,  0.988) & 78 \\
    $\sigma$ & 0.3 & 0.266 & 0.036 & ( 0.202,  0.342) & 9 & 0.262 & 0.035 & ( 0.196,  0.338) & 177 \\
    $\beta$ & 0.7 & 0.704 & 0.037 & ( 0.633,  0.776) & 3 & 0.732 & 0.037 & ( 0.661,  0.805) & 43 \\
    \hline
  \end{tabular*}
  \caption{True values, posterior means, posterior standard deviations, 95\% credible intervals, and inefficiency factors. $\beta = 0.7$.}
  \label{table:sim0_7_MCMC}
  \normalsize
\end{table}

\noindent
Furthermore, MCMC estimation results using Chan's method are shown in Tables \ref{table:chan0_3_MCMC}, \ref{table:chan0_5_MCMC} and \label{table:chan0_7_MCMC} for $\beta=0.3, 0.5$ and 0.7. Our proposed methods are found to be more efficient with respect to model parameters as in sampling $h_t$'s.
\begin{table}[H]
\small
  \centering
   \begin{tabular*}{\textwidth}{lr@{\extracolsep{\fill}}*{6}{r}}    \hline 
   \multicolumn{6}{c}{CHN $(\beta=0.3)$} \\
   \cline{3-6}
    Par & True & Mean & Std Dev & 95\% interval & IF \\
    \hline
    $\mu$ & 0 & 0.087 & 0.300 & (-0.516,  0.680) & 27 \\
    $\phi$ & 0.97 & 0.971 & 0.011 & ( 0.947,  0.989) & 174 \\
    $\sigma$ & 0.3 & 0.262 & 0.037 & ( 0.196,  0.341) & 249 \\
    $\beta$ & 0.3 & 0.328 & 0.033 & ( 0.263,  0.393) & 12 \\
    \hline
  \end{tabular*}
  \caption{True values, posterior means, posterior standard deviations, 95\% credible intervals, and inefficiency factors. $\beta = 0.3$.}
  \label{table:chan0_3_MCMC}
  \normalsize
\end{table}

\begin{table}[H]
\small
  \centering
   \begin{tabular*}{\textwidth}{lr@{\extracolsep{\fill}}*{6}{r}}    \hline 
   \multicolumn{6}{c}{CHN $(\beta=0.5)$} \\
   \cline{3-6}
    Par & True & Mean & Std Dev & 95\% interval & IF \\
    \hline
    $\mu$ & 0 & 0.088 & 0.319 & (-0.573,  0.726) & 18 \\
    $\phi$ & 0.97 & 0.972 & 0.011 & ( 0.949,  0.991) & 141 \\
    $\sigma$ & 0.3 & 0.258 & 0.036 & ( 0.192,  0.339) & 266 \\
    $\beta$ & 0.5 & 0.530 & 0.034 & ( 0.464,  0.600) & 23 \\
    \hline
  \end{tabular*}
  \caption{True values, posterior means, posterior standard deviations, 95\% credible intervals, and inefficiency factors. $\beta = 0.5$.}
  \label{table:chan0_5_MCMC}
  \normalsize
\end{table}

\begin{table}[H]
\small
  \centering
   \begin{tabular*}{\textwidth}{lr@{\extracolsep{\fill}}*{6}{r}}    \hline 
   \multicolumn{6}{c}{CHN $(\beta=0.7)$} \\
   \cline{3-6}
    Par & True & Mean & Std Dev & 95\% interval & IF \\
    \hline
    $\mu$ & 0 & 0.095 & 0.312 & (-0.527,  0.722) & 25 \\
    $\phi$ & 0.97 & 0.971 & 0.010 & ( 0.950,  0.988) & 173 \\
    $\sigma$ & 0.3 & 0.262 & 0.034 & ( 0.197,  0.331) & 277 \\
    $\beta$ & 0.7 & 0.730 & 0.037 & ( 0.658,  0.803) & 126 \\
    \hline
  \end{tabular*}
  \caption{True values, posterior means, posterior standard deviations, 95\% credible intervals, and inefficiency factors. $\beta = 0.7$.}
  \label{table:chan0_7_MCMC}
  \normalsize
\end{table}
\section{MCMC algorithm and particle filter for SVM with leverage}
\label{appendix:mcmc and particle filter leverage}
\subsection{MCMC algorithm}
 For the SVM with leverage, we set $\theta=(\mu,\phi,\sigma^2,\rho)$. For the prior distribution of $\rho$, we assume $\rho \sim U(-1,1)$ where $U(a,b)$ denotes uniform distribution over $(a,b)$, and the posterior density function of $(h, \theta)$ is given by
\begin{align*}
    \pi(&h, \theta|y) \\
    &\propto f(y, h|\theta) \pi(\theta) \\
    &\propto (1+\phi)^{a-\frac{1}{2}} (1-\phi)^{b-\frac{1}{2}} (1-\rho^2)^{-\frac{n-1}{2}} (\sigma^2)^{-\left( \frac{n_1}{2}+1 \right)} \exp \left\{ -\frac{1}{2\sigma^2} \{ s_0 + (1-\phi^2)(h_1-\mu)^2 \} \right\} \\
    &\quad \times \exp \left\{ -\frac{1}{2} \sum_{t=1}^n [ h_t + \{ y_t -\beta\exp(h_t/2) \}^2 \exp(-h_t) ] \right\} \\
    &\quad \times \exp \left\{ -\frac{1}{2\sigma^2(1-\rho^2)} \sum_{t=1}^{n-1} [ h_{t+1} - \mu(1-\phi) -\phi h_t - \rho \sigma \{ y_t -\beta\exp(h_t/2) \} \exp(-h_t/2) ]^2 \right\} \\
    &\times \exp \left\{ -\frac{(\mu-\mu_0)^2}{2\sigma_0^2} \right\} \exp \left\{ -\frac{(\beta-b_0)^2}{2B_0} \right\},
\end{align*}
where $n_1 = n_0 + n$.\\
\bigskip

\noindent
\textbf{Algorithm 3}.
Let us denote $\theta = (\alpha,\beta)$ where $\alpha = (\mu,\phi,\sigma^2,\rho)$. The Markov chain Monte Carlo simulation is implemented in four blocks:
\begin{itemize}
    \item[1.] Initialize $h$ and $\theta=(\alpha,\beta)$.
    \item[2.] Generate $\beta|\alpha,h, y \sim \pi(\beta|\alpha,h, y)$.
    \item[3.] Generate $(\alpha,h)|\beta, y \sim \pi(\alpha,h|\beta, y)$.
    \item[4.] Go to step 2.
\end{itemize}
\subsubsection*{Step 2. Generation of $\beta|\alpha,h,y$}
The conditional posterior distribution of $\beta$ is normal with mean $b_1$ and variance $B_1$ where
\begin{eqnarray*}
b_1 = B_1 \left(X'\Omega^{-1}\tilde{y} + B_0^{-1}b_0\right), \quad 
B_1^{-1} = X'\Omega^{-1}X + B_0^{-1},
\end{eqnarray*}
and
\begin{eqnarray*}
\tilde{y}&=&
\left(
	\begin{array}{c} 
	y_1 - \rho \exp(h_1/2)\sigma^{-1}\{h_2-\mu-\phi(h_1-\mu)\}\\
	y_2 - \rho \exp(h_2/2)\sigma^{-1}\{h_3-\mu-\phi(h_2-\mu)\}\\
        \vdots \\
	y_{n-1} - \rho \exp(h_{n-1}/2)\sigma^{-1}\{h_n-\mu-\phi(h_{n-1}-\mu)\}\\
 	y_n
	\end{array}  \right),
\quad
X = \left(\begin{array}{c}  \exp(h_1/2) \\  \exp(h_2/2) \\  \vdots \\ \exp(h_n/2)  \end{array}  \right),
\\
\Omega &=& \mbox{diag}\left((1-\rho^2)\exp(h_1),(1-\rho^2)\exp(h_2),\cdots,(1-\rho^2)\exp(h_{n-1}), \exp(h_n)\right),
\end{eqnarray*}
Thus we generate $\beta \sim N(b_1,B_1).$

\subsubsection*{Step 3. Generation of $(\alpha,h)|\beta,y$}
We define the pseudo target density
\begin{align*}
    \Tilde{\pi}(\alpha,h,s|\beta, y) &= \pi(\alpha,h|\beta,y) \times q(s|\alpha,h,\beta, y^*, d), \\
    &q(s|\alpha,h,\beta, y^*, d) = \prod_{t=1}^n \frac{\Tilde{p}_{s_t} g(y_t^*, h_{t+1}| h_t, \alpha,\beta, s_t, d)}{\sum_{i=1}^{10}\sum_{j=0}^2 \Tilde{p}_{i,j} g(y_t^*, h_{t+1}| h_t, \alpha,\beta, s_t=(i,j), d)}, 
\end{align*}
where $\theta=(\alpha,\beta)$ and
\begin{align*}
    g(y_t^*, &h_{t+1}| h_t, \alpha, \beta, s_t, d) = 
    \begin{cases}
        f_N(y_t^*|\Tilde{m}_{s_t} + h_t, v_{s_t}^2) f_N(h_{t+1}|\overline{h}_{s_t, t}, \sigma^2(1-\rho^2)), & t < n, \\
        f_N(y_t^*|\Tilde{m}_{s_t} + h_t, v_{s_t}^2) & t = n, 
    \end{cases} \\
    &\overline{h}_{s_t, t} = \mu(1-\phi) + \phi h_t +  \rho \sigma[d_t \exp(\Tilde{m}_{s_t}/2) \{ a_{s_t} + b_{s_t} (y_t^* - h_t - \Tilde{m}_{s_t})\} - \beta],
\end{align*}
where $y_t^* = \log(y_t^2)$, $d_t = I(y_t \geq 0) - I(y_t < 0)$.
$\Tilde{m}_{s_t} = \Tilde{m}_{s_{1t}, s_{2t}}$ and $\Tilde{p}_{s_t} = \Tilde{p}_{s_{1t}, s_{2t}}$ are defined in (\ref{approx_noncentral_chi2}) and $(p_{s_{1t}}, m_{s_{1t}}, v_{s_{1t}}^2, a_{s_{1t}}, b_{s_{1t}})$ are given in Table \ref{table:approx_chi2}.
Only $\Tilde{p}_{s_t}$, which depends on $\beta$, needs to be updated according to the formula in (\ref{approx_noncentral_chi2}) before sampling.
Note that the marginal density $\pi(\alpha,h|\beta,y)$ is our target density,  $\pi(\alpha,h|\beta,y) = \sum_s \Tilde{\pi}(\alpha,h,s|\beta,y)$. We generate sample $(\alpha, h, s)$ from the pseudo target density in two steps. 
\begin{itemize}
    \item[(a)] Generate $s \sim q(s|h, \theta, y^*, d)$.
    
    \item[(b)] Generate $(\alpha,h)|\theta, s, y \sim \Tilde{\pi}(\alpha, h|\beta, s, y)$.
           \begin{itemize}
        \item[(i)]  Generate $\alpha \sim \pi^*(\alpha|s,\beta,y^*,d)$.
        The target density here is given by
        \begin{align*}
        \pi^*(\alpha| s, \beta, y^*,d) \propto  m(y^*|\alpha,s,\beta,d) \pi(\alpha),
        \end{align*}
        where 
        \begin{align*}
        m(y^*|\alpha,s,\beta,d) = \int \prod_{t=1}^n g(y_t^*, h_{t+1}| h_t, \alpha, s_t, \beta, d) \times f_N \left( h_1 \bigg| \mu, \frac{\sigma^2}{1-\phi^2} \right) dh,
        \end{align*}
        which we evaluate using Kalman filter algorithm.
        We first transform $\alpha$ to $\vartheta = (\mu, \log\{ (1+\phi)/(1-\phi) \}, \log \sigma^2, \log\{ (1+\rho)/(1-\rho) \})$ to remove parameter constraints, and conduct MH algorithm to sample from the conditional posterior distribution with density $\pi^*(\vartheta|s,\beta,y) = \pi^*(\alpha|s, \beta,y) \times |d\alpha / d\vartheta|$ where $|d\alpha / d\vartheta|$ is the Jacobian of the transformation. 
        Compute the posterior mode $\hat{\vartheta}$ and define $\vartheta_*$ and $\Sigma_*$ as
        \begin{equation*}
        \vartheta_* = \hat{\vartheta}, \quad
        \Sigma_*^{-1} = -\frac{\partial^2 \log \pi^*(\vartheta|s,\beta, y)}{\partial \vartheta \partial \vartheta'} \bigg|_{\vartheta = \hat{\vartheta}}.
        \end{equation*}
    
        Given the current value $\vartheta$, generate a candidate $\vartheta^\dag$ from the distribution $N(\vartheta_*, \Sigma_*)$ and accept it with probability
        \begin{equation*}
        \alpha(\vartheta, \vartheta^\dag|s,\beta, y) = \min \left\{1, \frac{\pi^*(\vartheta^\dag|s, \beta, y) f_N(\vartheta|\vartheta_*, \Sigma_*)}{\pi^*(\vartheta|s, \beta,y) f_N(\vartheta^\dag|\vartheta_*, \Sigma_*)} \right\},
        \end{equation*}
        where $f_N(\cdot|\vartheta_*, \Sigma_*)$ is the probability density of $N(\vartheta_*, \Sigma_*)$. 
        If candidate $\vartheta^\dag$ is rejected, we take the current value $\vartheta$ as the next draw. When the Hessian matrix is not negative definite, we may take a flat normal proposal $N(\vartheta_*, c_0 I)$ using some large constant $c_0$. The obtained draw is denoted as $\alpha^\dag$.
        \item[(ii)] Generate $h|\alpha, s, \beta, y \sim \pi^*(h|\alpha, s, \beta, y)$. Given $\alpha=\alpha^\dag$, we propose a candidate $h^\dag = (h_1^\dag, ..., h_n^\dag)$ using a simulation smoother introduced by \cite{DeShephard(95)} and \cite{DurbinKoopman(02)} for the linear space Gaussian state space model as in (\ref{approx_linaer_start})-(\ref{approx_linaer_end}).
        The $h^\dag$ is a sample from
               \begin{equation*}
            \pi^*(h|\alpha^{\dagger}, s, \beta, y^*, d) = \frac{\prod_{t=1}^n g(y_t^*, h_{t+1}| h_t, \alpha^{\dagger},\beta, s_t, d)}{m(y^*|\alpha^{\dagger}, s,\beta,d)} \times f_N \left( h_1 \bigg| \mu^{\dagger}, \frac{\sigma^{2\dagger}}{1-\phi^{\dagger 2}} \right),
        \end{equation*}
        \item[(iii)] Generate $(\alpha,h)\sim \tilde{\pi}(\alpha,h|s,\beta,y^*,d)$. From (i) and (ii), we have a sample $(\alpha^\dagger,h^\dag)$ from $\pi^*(h|\alpha, s,\beta, y^*,d)\pi^*(\alpha|s,\beta,y^*,d).$
        Let
        \begin{align*}
            f(y_t, &h_{t+1}|h_t, \alpha,\beta) \\
            &= 
            \begin{cases}
                f_N(y_t| \beta \exp(h_t/2), \exp(h_t)) f_N(h_{t+1}| \overline{h}_t, \sigma^2(1-\phi^2)), & t < n \\
                f_N(y_t| \beta \exp(h_t/2), \exp(h_t)), & t = n,
            \end{cases} \\
            &\quad \overline{h}_t = \mu(1-\phi) + \phi h_t + \rho \sigma \{ y_t - \beta \exp(h_t/2) \} \exp(-h_t/2).
        \end{align*}
        Given the current value $(\alpha,h)$, accept the candidate $(\alpha^\dag,h^\dag)$ with probablity
        \begin{align*}
            \min &\left\{ 1, \frac{\Tilde{\pi}(\alpha^\dag, h^\dag| s, \beta, y)\pi^*(h|\alpha, s, \beta,y ^*,d)\pi^*(\alpha|\beta,s,y^*,d)}{\Tilde{\pi}(\alpha, h|s, \beta,y)\pi^*(h^\dag|\alpha^{\dagger}, s, \beta, y^*,d)\pi^*(\alpha^{\dagger}|\beta,s,y^*,d)} \right\} \\
            &= \min \left\{ 1, \frac{\pi(\alpha^\dag, h^\dag| \beta,y) q(s|h^\dag,\alpha^\dag,\beta, y^*) \pi^*(h|\alpha, s,\beta, y^*)\pi^*(\alpha|\beta,s,y^*,d)}{\pi(\alpha,h|\beta,y) q(s|h,\alpha,\beta, y^*) \pi^*(h^\dag|\alpha^{\dagger}, s, \beta, y^*)\pi^*(\alpha^{\dagger}|\beta,s,y^*,d)} \right\} \\
            &= \min \left\{ 1, \frac{q(s|h^\dag,\alpha^\dag,\beta, y^*) \prod_{t=1}^n f(y_t, h_{t+1}^\dag| h_t^\dag, \alpha^\dag,\beta) g(y_t^*, h_{t+1}| h_t, \alpha,\beta, s_t, d) }{q(s|h,\alpha,\beta, y^*)\prod_{t=1}^n f(y_t, h_{t+1}| h_t, \alpha,\beta) g(y_t^*, h_{t+1}^\dag| h_t^\dag, \alpha^\dag,\beta, s_t, d) } \right\} \\
            &= \min \left\{ 1, \frac{\prod_{t=1}^n f(y_t| h_t^\dag, \alpha^\dag,\beta) \sum_{i=1}^{10}\sum_{j=0}^2 \Tilde{p}_{i,j} g(y_t^*| h_t, \alpha,\beta, s_t=(i,j)) }{\prod_{t=1}^n f(y_t| h_t, \alpha,\beta) \sum_{i=1}^{10}\sum_{j=0}^2 \Tilde{p}_{i,j} g(y_t^*| h_t^\dag, \alpha^\dag,\beta, s_t=(i,j)) } \right\} \\
            &= \min \left\{ 1, \frac{\prod_{t=1}^n f(y_t, h_{t+1}^\dag| h_t^\dag, \alpha^\dag,\beta) \sum_{i=1}^{10}\sum_{j=0}^2 \Tilde{p}_{i,j} g(y_t^*, h_{t+1}| h_t, \alpha,\beta, s_t=(i,j), d) }{\prod_{t=1}^nf(y_t, h_{t+1}| h_t, \alpha,\beta) \sum_{i=1}^{10}\sum_{j=0}^2 \Tilde{p}_{i,j} g(y_t^*, h_{t+1}^\dag| h_t^\dag, \alpha^\dag,\beta, s_t=(i,j), d)) } \right\} \\
       \end{align*}
    \end{itemize}
\end{itemize}
\bigskip

\noindent
{\it Remark}. As in Algorithm 1, we may skip (iii) of Step 3b since the approximation error is usually small. 
\subsection{Associated particle filter}
We describe how to compute the likelihood $f(y|\theta)$ when there is a leverage effect. Let
\begin{align*}
    f(y_t|h_t, \theta) &= \frac{1}{\sqrt{2\pi}} \exp \left[ -\frac{1}{2}h_t - \frac{1}{2} \{ y_t - \beta \exp(h_t/2) \}^2 \exp(-h_t) \right] \\
    f(h_{t+1}|h_t, y_t, \theta) &= \frac{1}{\sqrt{2\pi (1-\rho^2)} \sigma} \exp \left\{ -\frac{(h_{t+1} - \mu_{t+1})^2}{2(1-\rho^2) \sigma^2} \right\}, \\
    \mu_{t+1} &= \mu + \phi(h_t-\mu) + \rho \sigma \exp(-h_t/2) \{ y_t - \beta \exp(h_t/2) \},
\end{align*}
and consider the importance function for the auxiliary particle filter
\begin{align*}
    q(h_{t+1}, h_t^i| Y_{t+1}, \theta) &\propto  f(y_{t+1}| \mu_{t+1}^i, \theta) f(h_{t+1}|h_t^i, y_t, \theta) \hat{f}(h_t^i|Y_t, \theta) \\
    &\propto f(h_{t+1}|h_t^i, y_t, \theta) q(h_t^i| Y_{t+1}, \theta)
\end{align*}
where
\begin{align*}
    q(h_t^i| Y_{t+1}, \theta) &= \frac{f(y_{t+1}| \mu_{t+1}^i, \theta) \hat{f}(h_t^i|Y_t, \theta)}{\sum_{j=1}^I f(y_{t+1}| \mu_{t+1}^j, \theta) \hat{f}(h_t^j|Y_t, \theta)}, \\
    f(y_{t+1}|\mu_{t+1}^i, \theta) &= \frac{1}{\sqrt{2\pi}} \exp \left[ -\frac{1}{2}\mu_{t+1}^i - \frac{1}{2} \{ y_t - \beta \exp(h_t^i/2) \}^2 \exp(-\mu_{t+1}^i) \right], \\
    \mu_{t+1}^i &= \mu + \phi(h_t^i-\mu) + \rho \sigma \exp(-h_t^i/2) \{ y_t - \beta \exp(h_t^i/2) \}.    
\end{align*}
This leads to the following particle filtering.
\begin{itemize}
    \item[1.] Compute $\hat{f}(y_1|\theta)$ and $\hat{f}(h_1^i| Y_1, \theta) = \pi_1^i$ for $i = 1, \dots, I$. 
    
    \begin{itemize}
        \item[(a)] Generate $h_1^i \sim f(h_1|\theta)$ ($= N(\mu, \sigma^2/(1-\phi^2))$) for $i = 1, \dots, I$.
        
        \item[(b)] Compute
        \begin{align*}
            &\pi_1^i = \frac{w_i}{\sum_{i=1}^I w_i}, \quad
            w_i = f(y_1|h_1, \theta), \quad 
            W_i = F(y_1|h_1, \theta), \\
            &\hat{f}(y_1|\theta) = \overline{w}_1 = \frac{1}{I} \sum_{i=1}^I w_i, \quad
            \hat{F}(y_1|\theta) = \overline{W}_1 = \frac{1}{I} \sum_{i=1}^I W_i,
        \end{align*}
        where $f(y_1|\theta)$ and $F(y_1|\theta)$ are the marginal density function and the marginal distribution function of $y_1$ given $\theta$. Let $t = 1$.
    \end{itemize}

    \item[2.] Compute $\hat{f}(y_{t+1}|\theta)$ and $\hat{f}(h_{t+1}^i| Y_{t+1}, \theta) = \pi_{t+1}^i$ for $i = 1, \dots, I$.
    
    \begin{itemize}
        \item[(a)] Sample $h_t^i \sim q(h_t| Y_t, \theta)$, $i = 1, \dots, I$.

        \item[(b)] Generate $h_{t+1}^i| h_t^i, y_t, \theta \sim f(h_{t+1}| h_t^i, y_t, \theta)$ ($= N(\mu_{t+1}^i, \sigma^2(1-\rho^2))$) for $i = 1, \dots, I$.  

        \item[(c)] Compute
    \end{itemize}
       \begin{align*}
            &\pi_{t+1}^i = \frac{w_i}{\sum_{i=1}^I w_i}, \quad
            w_i = \frac{ f(y_{t+1}|h_{t+1}^i, \theta) f(h_{t+1}^i| h_t^i, y_t, \theta) \hat{f}(h_t^i| Y_t, \theta) }{ f(h_{t+1}^i| h_t^i, y_t, \theta) q(h_t^i| Y_{t+1}, \theta) } = \frac{ f(y_{t+1}|h_{t+1}^i, \theta) \hat{f}(h_t^i| Y_t, \theta) }{ q(h_t^i| Y_{t+1}, \theta) }, \\ 
            &W_i = \frac{ F(y_{t+1}|h_{t+1}^i, \theta) \hat{f}(h_t^i| Y_t, \theta) }{ q(h_t^i| Y_{t+1}, \theta) }, \\
            &\hat{f}(y_{t+1}| Y_t, \theta) = \overline{w}_{t+1} = \frac{1}{I} \sum_{i=1}^I w_i, \quad
            \hat{F}(y_{t+1}|\theta) = \overline{W}_{t+1} = \frac{1}{I} \sum_{i=1}^I W_i.
        \end{align*}    

        \item[3.] Increment $t$ and go to 2.
\end{itemize}

\section{MCMC algorithm to compute the posterior ordinate}
\label{appendix:posterior ordinate}
This section describes MCMC algorithm which may be used when computing the marginal likelihood. It is a little less efficient than Algorithm 3, but still efficient enough to compute the posterior ordinate.\\

\noindent
{\bf Algorithm 4}.
The Markov chain Monte Carlo simulation is implemented in four blocks:

\begin{itemize}
    \item[1.] Initialize $h$ and $\theta$.
    \item[2.] Generate $\theta|h, y \sim \pi(\theta|h, y)$.
    \item[3.] Generate $h|\theta, y \sim \pi(h|\theta, y)$.
    \item[4.] Go to step 2.
\end{itemize}
\subsubsection*{Step 2. Generation of $\theta|h,y$}
We first transform $\theta$ to $\vartheta = (\mu, \log\{ (1+\phi)/(1-\phi) \}, \log \sigma^2, \beta, \log\{ (1+\rho)/(1-\rho) \}), $ to remove parameter constraints, and conduct Metropolis-Hastings (MH) algorithm to sample from the conditional posterior distribution with density $\pi(\vartheta|h, y) = \pi(\theta|h, y) \times |d\theta / d\vartheta|$ where $|d\theta / d\vartheta|$ is the Jacobian of the transformation. 
Compute the posterior mode $\hat{\vartheta}$ and define $\vartheta_*$ and $\Sigma_*$ as
\begin{equation*}
    \vartheta_* = \hat{\vartheta}, \quad
    \Sigma_*^{-1} = -\frac{\partial^2 \log \pi(\vartheta|h, y)}{\partial \vartheta \partial \vartheta'} \bigg|_{\vartheta = \hat{\vartheta}}.
\end{equation*}
Given the current value $\vartheta$, generate a candidate $\vartheta^\dag$ from the distribution $N(\vartheta_*, \Sigma_*)$ and accept it with probability
\begin{equation*}
    \alpha(\vartheta, \vartheta^\dag|h, y) = \min \left\{1, \frac{\pi(\vartheta^\dag|h, y) f_N(\vartheta|\vartheta_*, \Sigma_*)}{\pi(\vartheta|h, y) f_N(\vartheta^\dag|\vartheta_*, \Sigma_*)} \right\},
\end{equation*}
where $f_N(\cdot|\vartheta_*, \Sigma_*)$ is the probability density of $N(\vartheta_*, \Sigma_*)$.
If candidate $\vartheta^\dag$ is rejected, we take the current value $\vartheta$ as the next draw. When the Hessian matrix is not negative definite, we may take a flat normal proposal $N(\vartheta_*, c_0 I)$ using some large constant $c_0$.
\subsubsection*{Step 3. Generation of $h|\theta,y$}
We sample $h$ using the mixture sampler using the mixture of normal distributions as discussed in the previous section. Define the pseudo target density
\begin{align*}
    \Tilde{\pi}(h,s|\theta,y) &= \pi(h|\theta,y) \times q(s|h,\theta, y^*, d), \\
    &q(s|h,\theta, y^*, d) = \prod_{t=1}^n \frac{\Tilde{p}_{s_t} g(y_t^*, h_{t+1}| h_t, \theta, s_t, d)}{\sum_{i=1}^{10}\sum_{j=0}^2 \Tilde{p}_{i,j} g(y_t^*, h_{t+1}| h_t, \theta, s_t=(i,j), d)}, 
\end{align*}
where
\begin{align*}
    g(y_t^*, &h_{t+1}| h_t, \theta, s_t, d) = 
    \begin{cases}
        f_N(y_t^*|\Tilde{m}_{s_t} + h_t, v_{s_t}^2) f_N(h_{t+1}|\overline{h}_{s_t, t}, \sigma^2(1-\rho^2)), & t < n, \\
        f_N(y_t^*|\Tilde{m}_{s_t} + h_t, v_{s_t}^2) & t = n, 
    \end{cases} \\
    &\overline{h}_{s_t, t} = \mu(1+\phi) + \phi h_t +  \rho \sigma[d_t \exp(\Tilde{m}_{s_t}/2) \{ a_{s_t} + b_{s_t} (y_t^* - h_t - \Tilde{m}_{s_t})\} - \beta].
\end{align*}
We generate sample $(h, s)$ from the pseudo target density in two steps. 
\begin{itemize}
    \item[(a)] Generate $s \sim q(s|h, \theta, y^*, d)$.
    
    \item[(b)] Generate $h|\theta, s, y \sim \Tilde{\pi}(h|\theta, s, y)$.
    
    \begin{itemize}
        \item[i.] Propose a candidate $h^\dag = (h_1^\dag, ..., h_n^\dag)$ using a simulation smoother introduced by \cite{DeShephard(95)} and \cite{DurbinKoopman(02)} for the linear space Gaussian state space model as in (\ref{approx_linaer_start})-(\ref{approx_linaer_end}).
        The $h^\dag$ is a sample from
        \begin{equation*}
            \pi^*(h|\theta, s, y^*, d) = \frac{\prod_{t=1}^n g(y_t^*, h_{t+1}| h_t, \theta, s_t, d)}{m(y^*|\theta, s)} \times f_N \left( h_1 \bigg| \mu, \frac{\sigma^2}{1-\phi^2} \right),
        \end{equation*}
        where $m(y^*|\theta, s)$ is a normalizing constant given by 
        \begin{equation*}
            m(y^*|\theta, s) = \int \prod_{t=1}^n g(y_t^*, h_{t+1}| h_t, \theta, s_t, d) \times f_N \left( h_1 \bigg| \mu, \frac{\sigma^2}{1-\phi^2} \right) dh.
        \end{equation*}

        \item[ii.] Let
        \begin{align*}
            f(y_t, &h_{t+1}|h_t, \theta) \\
            &= 
            \begin{cases}
                f_N(y_t| \beta \exp(h_t/2), \exp(h_t)) f_N(h_{t+1}| \overline{h}_t, \sigma^2(1-\rho^2)), & t < n \\
                f_N(y_t| \beta \exp(h_t/2), \exp(h_t)), & t = n,
            \end{cases} \\
            &\quad \overline{h}_t = \mu(1+\phi) + \phi h_t + \rho \sigma \{ y_t - \beta \exp(h_t/2) \} \exp(-h_t/2).
        \end{align*}
        Given the current value $h$, accept the candidate $h^\dag$ with probability
        \begin{align*}
            \min &\left\{ 1, \frac{\Tilde{\pi}(h^\dag|\theta, s, y)\pi^*(h|\theta, s, y^*, d)}{\Tilde{\pi}(h|\theta, s, y)\pi^*(h^\dag|\theta, s, y^*, d)} \right\} \\
            &= \min \left\{ 1, \frac{\pi(h^\dag|\theta,y) q(s|h^\dag,\theta, y^*, d) \pi^*(h|\theta, s, y^*, d)}{\pi(h|\theta,y) q(s|h,\theta, y^*, d) \pi^*(h^\dag|\theta, s, y^*, d)} \right\} \\
            &= \min \left\{ 1, \frac{ q(s|h^\dag,\theta, y^*, d) \prod_{t=1}^n f(y_t, h_{t+1}^\dag| h_t^\dag, \theta) g(y_t^*, h_{t+1}| h_t, \theta, s_t, d) }{ q(s|h,\theta, y^*, d) \prod_{t=1}^n f(y_t, h_{t+1}| h_t, \theta) g(y_t^*, h_{t+1}^\dag| h_t^\dag, \theta, s_t, d) } \right\} \\
            &= \min \left\{ 1, \frac{ \prod_{t=1}^n f(y_t, h_{t+1}^\dag| h_t^\dag, \theta) \sum_{i=1}^{10}\sum_{j=0}^2 \Tilde{p}_{i,j} g(y_t^*, h_{t+1}| h_t, \theta, s_t=(i,j), d)) }{ \prod_{t=1}^n f(y_t, h_{t+1}| h_t, \theta) \sum_{i=1}^{10}\sum_{j=0}^2 \Tilde{p}_{i,j} g(y_t^*, h_{t+1}^\dag| h_t^\dag, \theta, s_t=(i,j), d)) } \right\}. \\ 
        \end{align*}
    \end{itemize}
\end{itemize}

\end{document}